\documentclass[12pt]{article}
\usepackage{epsfig,amsmath,amssymb,amsthm,thmtools,latexsym,graphicx,setspace,fullpage}
\usepackage{chicago} 
\usepackage{rotating}
\usepackage{breqn}
\usepackage{authblk}

\allowdisplaybreaks
\usepackage{color} 

\declaretheoremstyle[
notefont=\bfseries, notebraces={}{},
bodyfont=\normalfont,
postheadspace=0.5em,
numbered=no,
]{mystyle}

\theoremstyle{definition}
\newtheorem{exmp}{Example}[section]

\begin{document}

\renewcommand{\thefootnote}{\fnsymbol{footnote}}

\title{\vspace{-5mm}Checking for prior-data conflict using prior \\ to posterior divergences}

\author[1,2]{David J. Nott\thanks{Corresponding author:  standj@nus.edu.sg}}
\author[1]{Wang Xueou}
\author[3]{Michael Evans}
\author[4,5,6]{Berthold-Georg Englert}
\affil[1]{Department of Statistics and Applied Probability, National University of Singapore, Singapore 117546}
\affil[2]{Operations Research and Analytics Cluster,
National University of Singapore, Singapore 119077}
\affil[3]{Department of Statistics, University of Toronto, Toronto, Ontario, M5S 3G3, Canada}
\affil[4]{Department of Physics,
National University of Singapore, 
Singapore 117542}
\affil[5]{Centre for Quantum Technologies, National University of Singapore, 3 Science Drive 2, Singapore 117543}
\affil[6]{MajuLab, CNRS-UNS-NUS-NTU International Joint Research Unit, UMI 3654, Singapore.}

\date{}

\maketitle

\onehalfspacing

\vspace*{-11mm}\noindent

\begin{abstract}
\noindent  
When using complex Bayesian models to combine information, the checking for
consistency of the information being combined is good statistical practice.  
Here a new method is developed for detecting prior-data conflicts in Bayesian
models based on comparing the observed
value of a prior to posterior divergence to its distribution under the prior predictive distribution for the data.  
The divergence measure used in our model check is a measure of how much beliefs have changed from prior
to posterior, and can be thought of as a measure of the overall size of a relative belief
function. 
It is shown that the proposed method is intuitive, has desirable properties, can be extended to hierarchical settings, and 
is related asymptotically to Jeffreys' and reference prior distributions.  
In the case where calculations are difficult, the use of variational approximations
as a way of relieving the computational burden is suggested.  
The methods are compared in a number of examples with an alternative but closely related approach in the literature based on the 
prior predictive distribution of a minimal sufficient statistic.  

\vspace{2mm}

\noindent{\bf Keywords}:  Bayesian inference, Model checking, Prior data-conflict, Variational Bayes.
\end{abstract}

\section{Introduction}

In modern applications, statisticians are often confronted with the task of 
either combining data and expert knowledge, or of combining information
from diverse data sources using hierarchical models.  In these settings, Bayesian methods are very useful.  
However, whenever we perform Bayesian inference combining different sources of information,
it is important to check the consistency of the information being combined.  
This work is concerned with the problem of detecting situations in which information coming from the prior and the data 
are in conflict in a Bayesian analysis.  
Such conflicts can highlight a lack of understanding of the information put into the model, and
it is only when there is no conflict between prior and data that we can expect Bayesian inferences to show robustness
to the prior \cite{allabadi+e15}.  See \shortciteN{andrade+o06} for a discussion of Bayesian robustness and the behaviour of Bayesian inferences in the
case of prior-data conflict.  

Here a new and attractive approach to measuring prior-data conflict is introduced based on a prior to posterior divergence, and
the comparison of the observed value of this statistic with its prior predictive distribution.  We show that this method 
extends easily to hierarchical settings, and has an interesting relationship asymptotically with Jeffreys' and reference prior distributions.  For the prior to posterior divergence, we consider
the class of R\'{e}nyi divergences \cite{renyi61}, with the Kullback-Leibler divergence as an important special case.  In the present context, the R\'{e}nyi divergence can be thought
of as giving an overall measure of the size of a relative belief function, which is a function describing for each possible value of a given parameter of interest how much more or less likely it
has become after observing the data.  
\shortciteN{evans15} and \shortciteN{baskurt+e13} give details 
of some attractive solutions to many inferential problems based on the notion of relative belief.  
A large change in beliefs from prior to posterior (where this is calibrated by the prior predictive) may be indicative of conflict between prior and likelihood,
so that a check
with prior to posterior R\'{e}nyi divergence as the checking discrepancy is an intuitive one for prior-data conflict detection.  

Checks for prior-data conflict have usually been formulated within the broader framework of Bayesian predictive model checking, although
much of this work is concerned with approaches which check the prior and model jointly
(see, for example, \shortciteN{gelman+ms96} and \shortciteN{bayarri+c07} 
for entries into this literature).  In general the idea is that there is a discrepancy function $D(y)$ of data $y$ (where a large value of this discrepancy might
represent an unusual value) and then for some reference predictive density $m(y)$ a $p$-value is computed as
\begin{align}
  p & = P\Bigl(D(Y)\geq D(y_{\text{obs}})\Bigr),  \label{predcheck}
\end{align}
where $Y\sim m(y)$ is a draw from the reference predictive distribution and $y_{\text{obs}}$ is the observed data.  A small $p$-value indicates that the 
observed value of the discrepancy is surprising under the assumed model, and that the model formulation might need to be re-examined.   
The choice of discrepancy will reflect some aspect of the model fit that we wish to check, and this is generally application specific.   The reference
predictive density $m(y)$ needs to be chosen, and there are many ways that this can be done.  
For example, $m(y)$ might be the prior predictive density $\int g(\theta)p(y|\theta) d\theta$ \cite{box80},
where $g(\theta)$ is the prior density and $p(y|\theta)$ is the density of $y$ given $\theta$.  Another common choice of 
reference distribution is the posterior predictive for
a hypothetical replicate \cite{guttman67,rubin84,gelman+ms96}.  More complex kinds of replication can also be considered, 
particularly in the case of hierarchical
models.  In some cases, the discrepancy might also be allowed to depend on the parameters, in which case the reference distribution defines a joint distribution on
both the parameters and $y$.  
When the discrepancy is chosen in a casual way in the posterior predictive approach it may be hard to interpret checks in a similar way across
different problems, and a variety of authors
have suggested modifications which have better calibration properties \cite{bayarri+b00,robins+vv00,hjort+ds06}.  
The choice of a suitable discrepancy and reference distribution in Bayesian predictive model checking often depends on statistical goals, 
and this is discussed more later.

Checking for prior-data conflict is distinct
from the issue of whether the likelihood component of the model is adequately specified.  
An incorrect likelihood specification means that there are no parameter values which provide a good fit to the data, whereas a prior-data conflict occurs
when the prior puts all its mass in the tails of the likelihood.  
See Chapter 5 of \shortciteN{evans15} for a discussion of
different kinds of model checks.  
Although we focus here on prior-data conflict checks, and not on checking the adequacy of the likelihood specification, 
\shortciteN{carota+pp96} describe one method for the latter problem related to the current work.  
They consider checking model adequacy by defining a model expansion and 
then measuring the utility of the
expansion.  Their preferred measure of utility is the marginal prior to posterior Kullback-Leibler divergence for the expansion parameter, 
and they consider calibration by comparison of the Kullback-Leibler divergence with its value in some reference situations involving
simple distributions.  Their use of a prior to posterior divergence in a model check is related
to our approach and an interesting complement to our method for prior-data conflict checking.  The approach is very flexible, but the elements of their construction 
need to be chosen with care to avoid confounding prior-data conflict checking with assessing the adequacy of the likelihood, and their approach to calibration of the diagnostic
measure is also quite different.  

Henceforth we will focus exclusively on model checking with the aim of detecting prior-data conflicts.    
We postpone a comprehensive survey of the literature on prior-data conflict assessment to the next section, after first describing the basic idea of our own approach.  
However, one feature of many
existing suggestions for prior-data conflict checking is that they require the definition of a non-informative prior.  
Among methods that don't require such a choice our approach is closely related
to that of \shortciteN{evans+m06}. They modify the approach to model checking given by \shortciteN{box80} 
by considering as the checking discrepancy the prior predictive density value for a sufficient statistic, and they use the prior predictive distribution as the
reference predictive distribution.  They show that these choices are logical ones for the specific purpose of checking for prior-data conflict.  
We will use this method as a reference for comparison in our later examples.

In Section 2 we introduce the basic idea of our method and discuss its relationship with other approaches in the literature.  
In Section 3 a series of simple examples where calculations can be done analytically is described.  
In Section 4 we consider the asymptotic behaviour of the checks, 
and some more complex examples are considered in Section 5 where computational implementation using variational approximation methods is considered.  Section 6 concludes with
some discussion.  

\section{Prior-data conflict checking}

\subsection{The basic idea and relationship with relative belief}

Let $\theta$ be a $d$-dimensional parameter and $y$ be data to be observed.  
We will assume henceforth that all distributions such as the joint distribution for $(y,\theta)$ 
can be defined in terms of densities with respect to appropriate support measures and that in
the continuous case these densities are defined uniquely in terms of limits (see, for example, Appendix A of \shortciteN{evans15}).  
We consider Bayesian inference where the prior
density is $g(\theta)$ and $p(y|\theta)$ is the density of $y$ given $\theta$.  The posterior density is 
$g(\theta|y)\propto g(\theta)p(y|\theta)$.  
We consider checks for prior-data conflict based on a prior to posterior 
R\'{e}nyi divergence of order $\alpha$ \cite{renyi61} (sometimes referred to as an $\alpha$ divergence).  
\begin{align}
 R_\alpha(y) & =\frac{1}{\alpha-1} \log \int \left\{\frac{g(\theta|y)}{g(\theta)}\right\}^{\alpha-1} g(\theta|y) d\theta,  \label{renyi}
\end{align}
where $\alpha>0$ and the case $\alpha=1$ is defined by letting $\alpha\rightarrow 1$.  This
corresponds to the Kullback-Leibler divergence, and we write
$$\text{KL}(y)=\lim_{\alpha\rightarrow 1}R_\alpha(y)=\int \log \frac{g(\theta|y)}{g(\theta)} g(\theta|y)\;d\theta.$$
Also of interest is to consider $\alpha\rightarrow\infty$, which gives the maximum value of $\log \frac{g(\theta|y)}{g(\theta)}$, 
and we write$\text{MR}(y)=\lim_{\alpha\rightarrow\infty} R_\alpha(y)$.
Our proposed $p$-value for the prior-data conflict check is 
\begin{align}
  p_{\alpha}=p_\alpha(y_{\text{obs}})=P(R_\alpha(Y)\geq R_\alpha(y_{\text{obs}})) \label{klcheck}
\end{align}
where $y_{\text{obs}}$ is the observed value of $y$ and $Y\sim p(y)=\int g(\theta)p(y|\theta) d\theta$ is a draw from the prior predictive distribution.  
This is a measure of how surprising the observed value $R_\alpha(y_{\text{obs}})$ is in terms of its prior distribution.  For if this is small then 
the distance between the prior and posterior is much greater than expected.  
The use of $p$-values in Bayesian model checking as measures of surprise is well established, but we emphasize here that these $p$-values are not measures
of evidence, and it may be better to think of the tail probability (\ref{klcheck}) as a calibration of the observed value of 
$R_\alpha(y_{\text{obs}})$.  However, we will continue to use the well-established $p$-value terminology in what follows.    
We will use the special notation $p_{\text{KL}}$ and $p_{\text{MR}}$ for 
the $p$-values based on the discrepancies $\text{KL}(y)$ and $\text{MR}(y)$ respectively.  In the definition (\ref{renyi}) it was assumed that we want an overall conflict check
for the prior.  If interest centres on a particular quantity $\Psi(\theta)$, however, we can look at the marginal prior to posterior divergence for $\Psi$ instead of $\theta$ in
(\ref{renyi}).  

The prior-data conflict check (\ref{klcheck}) can be motivated from a number of points of view.  First, the choice of discrepancy is intuitive, since $R_\alpha(y)$ is a measure of
how much beliefs change from prior to posterior, and comparing this measure for $y_{\text{obs}}$ against what is expected under the prior predictive intuitively tells us something
about how surprising the observed data and likelihood are under the prior.  This point of view connects with the relative belief framework for inferences
summarized in \shortciteN{baskurt+e13} and \shortciteN{evans15}.  For a parameter of interest $\Psi=\Psi(\theta)$, the relative belief function is
the ratio of the posterior density of $\Psi$ to its prior density, 
\begin{align*}
  \text{RB}(\Psi|y) & = \frac{g(\Psi|y)}{g(\Psi)}.
\end{align*}
$\text{RB}(\Psi|y)$ measures how much belief in $\psi$ being the true value has changed after observing data $y$.  If $\text{RB}(\Psi|y)$ is bigger than $1$, this says that 
there is evidence for $\Psi$ being the true value, whereas if it is less than $1$ this says that there is evidence against.  
Use of the R\'{e}nyi divergence as the discrepancy in (\ref{klcheck}) is equivalent to the use of the discrepancy
\begin{align}
 \|\text{RB}(\theta|y)\|_s  & =E\Bigl(\text{RB}(\theta|y)^s|y\Bigr)^{1/s} \label{mrbstat}
\end{align}
as a test statistic, where $s=\alpha-1$, since $R_\alpha(y)=\log \|\text{RB}(\theta|y)\|_s$.  (\ref{mrbstat}) is a measure of the overall size of the relative belief function.  The limit $s\rightarrow 0$ gives
$\exp(\text{KL}(y))$, $s\rightarrow\infty$ gives $\text{RB}(\hat{\theta}|y)$ where $\hat{\theta}$ denotes the maximum relative belief estimate which maximizes the relative
belief function, and $s=1$ is the posterior mean of the relative belief.  

In Section 4 we also investigate the asymptotic behaviour of $p_{\alpha}$, which under appropriate conditions converges in the large data limit to
\begin{align}
  & P\Bigl(g(\theta^*)|I(\theta^*)|^{-1/2} \geq g(\theta)|I(\theta)|^{-1/2}\Bigr)  \label{klchecklimit}
\end{align}
where $I(\theta)$ is the Fisher information at $\theta$, $\theta^*$ is the true value of the parameter that generated the data, and $\theta\sim g(\theta)$.  
To interpret (\ref{klchecklimit}), note that $g(\theta)|I(\theta)|^{-1/2}$ is just the prior density, but written with respect to the Jeffreys' prior as the support measure rather
than Lebesgue measure.  So (\ref{klchecklimit}) is the probability that a draw from the prior has prior density value less than the prior density value at the true parameter.  
It is a measure of how far out in the tails of the prior the true value $\theta^*$ lies.  There is a similar limit result for the check of \shortciteN{evans+m06}, but where the densities
are with respect to Lebesgue measure \cite{evans+j11b}.  
Interestingly, (\ref{klchecklimit}) might be thought of as giving some kind of heuristic justification for why the Jeffreys' prior could be considered non-informative -- if we
were to choose $g(\theta)$ as the Jeffreys' prior, $g(\theta)\propto |I(\theta)|^{1/2}$  then the value of the 
limiting $p$-value (\ref{klchecklimit}) is $1$ and hence there can be no conflict asymptotically.
Some similar connections with reference priors
\cite{berger+bs09,ghosh11} are considered in Section 4 for hierarchical versions of our checks and we discuss these in Section 2.2.

Further motivation for the approach follows from some logical principles that any prior-data conflict check
should satisfy.  \shortciteN{evans+m06} and \shortciteN{evans+j11} consider for a minimal sufficient statistic $T$ a decomposition of the joint model as
\begin{align}
  p(\theta,y) & = p(t)g(\theta|t)p(y|\theta,t)=p(t)g(\theta|t)p(y|t)  \label{decomposition}
\end{align}
where the terms in the decomposition are densities with respect to appropriate support measures, $p(t)$ is the prior predictive density for
$T$, $g(\theta|t)$ is the density of $\theta$ given $T=t$ (which is the posterior density since $T$ is sufficent) 
and $p(y|t)$ is the density of $y$ given $T=t$ (which does not depend on $\theta$ because
of the sufficiency of $T$).  This decomposition generalizes a suggestion of \shortciteN{box80}.  
In the case where there is no non-trivial minimal sufficient statistic a decomposition (\ref{decomposition}) can still be contemplated for
some asymptotically sufficient $T$ such as the maximum likelihood estimator.  
The three terms in the decomposition could logically be specified separately in defining a joint model and they perform different
roles in an analysis.  For example, the posterior distribution $p(\theta|t)$ is used for inference, and $p(y|t)$ is useful for checking the likelihood, 
since it does not depend on the prior.  
Ideally a check of adequacy for the likelihood should not depend on the prior since the adequacy of the likelihood has nothing to do with the prior.  

For checking for prior-data conflict, \shortciteN{evans+m06} and \shortciteN{evans+j11} argue that the relevant part of the decomposition (\ref{decomposition}) 
is the prior predictive distribution of $T$.
Since a sufficient statistic determines the likelihood, a comparison between the likelihood
and prior can be done by comparing the observed value of a sufficient statistic to its prior predictive distribution.  Clearly any variation in $y$ that is not a function of a sufficient
statistic does not change the likelihood, and hence is irrelevant to determining whether prior and likelihood conflict.  Furthermore, a minimal sufficient statistic
will be best for excluding as much irrelevant variation as possible.  For a minimal sufficient statistic $T$, the $p$-value for the check of \shortciteN{evans+m06} is computed as
\begin{align}
  p_{\text{EM}}=p_{\text{EM}}(y_{\text{obs}})=P\Bigl(p(T)\leq p(t_{\text{obs}})\Bigr)  \label{emcheck}
\end{align}
where $t_{\text{obs}}$ is the observed value of $T$ and $T\sim p(t)$ is a draw from the prior predictive for $T$.
This approach, however, does not achieve invariance to the choice of the minimal sufficient statistic, which is generally not unique; see, however, \shortciteN{evans+j10} for
an alternative approach which does achieve invariance.  They also consider conditioning on maximal ancillary statistics when they are available.  
Coming back from these general principles to the check (\ref{klcheck}), 
we notice that the statistic $R_\alpha(y)$ is automatically a function of any sufficient statistic, since it depends on the data only through the posterior distribution.  Furthermore, it is the
same function no matter what sufficient statistic is chosen.  So our check is a function of any minimal sufficient statistic as \shortciteN{evans+m06} and \shortciteN{evans+j11}
would require, and is invariant to the particular choice of that statistic.

\subsection{Hierarchical versions of the check}

Next, consider implementation of the approach of Section 2.1 in a hierarchical setting.
Suppose the parameter $\theta$ is partitioned as $\theta=(\theta_1,\theta_2)$, where $\theta_1$ and $\theta_2$ are of dimensions $d_1$ and $d_2$ respectively, and that
the prior is decomposed as $g(\theta)=g(\theta_1|\theta_2)g(\theta_2)$ .  Sometimes it is natural to consider the decomposition of the prior into marginal 
and conditional pieces since it may reflect how the prior is specified (such as in the case of a hierarchical model).  
We may wish to check the two pieces of the prior separately to understand the nature of any prior-data conflict when it occurs.  
Mirroring our decomposition of the prior, write $g(\theta|y)=g(\theta_1|\theta_2,y)g(\theta_2|y)$.
To define a hierarchically structured check, let
\begin{align}
 R_\alpha(y,\theta_2) & =\frac{1}{\alpha-1}\log \int \left\{\frac{g(\theta_1|\theta_2,y)}{g(\theta_1|\theta_2)}\right\}^{\alpha-1}g(\theta_1|\theta_2,y)d\theta_1 \label{ralphay2}
\end{align}
denote the conditional prior to conditional posterior R\'{e}nyi divergence of order $\alpha$ for $\theta_1$ given $\theta_2$, and define
\begin{align}
  R_{\alpha 1}(y)=E_{\theta_2|y_{\text{obs}}}\Bigl(R_\alpha(y,\theta_2)\Bigr).  \label{Ralpha1}
\end{align}
$R_{\alpha 1}(y)$ is a function of both $y$ and $y_{\text{obs}}$ although we suppress this in the notation.  
Also, define
\begin{align*}
  R_{\alpha 2}(y)=\frac{1}{\alpha-1}\log \int \left\{\frac{g(\theta_2|y)}{g(\theta_2)}\right\}^{\alpha-1} g(\theta_2|y)d\theta_2
\end{align*}
so that $R_{\alpha 2}(y)$ is the marginal prior to posterior divergence for $\theta_2$.

For hierarchical checking of the prior we consider the $p$-values 
\begin{align}
  p_{\alpha 1}=P\Bigl(R_{\alpha 1}(Y)\geq R_{\alpha 1}(y_{\text{obs}})\Bigr)  \label{klcheck1}
\end{align}
where 
\begin{align}
Y\sim m(y)= & \int g(\theta_2|y_{\text{obs}})p(y|\theta)p(\theta_1|\theta_2)\;d\theta \label{refklcheck1} 
\end{align}
and
\begin{align}
  p_{\alpha 2} = & P\Bigl(R_{\alpha 2}(Y)\geq R_{\alpha 2}(y_{\text{obs}})\Bigr)  \label{klcheck2}
\end{align}
where $Y\sim p(y)=\int p(\theta)p(y|\theta)$.  The $p$-value (\ref{klcheck1}) is just measuring whether the conditional prior to posterior divergence for $\theta_1$
given $\theta_2$ is unusually large for values of $\theta_2$ and a reference distribution for $Y$ that reflects knowledge of $\theta_2$ under $y_{\text{obs}}$.   
The $p$-value (\ref{klcheck2}) is just the non-hierarchical check (\ref{klcheck}) applied to the marginal posterior and prior for $\theta_2$.  
We explore the behaviour of these hierarchical checks in examples later, as well as by examining their asymptotic behaviour in Section 4, where we find 
that these checks are related to two stage reference priors.  
In the above discussion we can also consider a partition of the parameters with more than two pieces and the ideas discussed can be extended without
difficulty to this more general case.  We can also consider functions of $\theta_1$ and $\theta_2$, $\Psi_1(\theta_1)$ and $\Psi_2(\theta_2)$, and prior to posterior
divergences involving these quantities in the definition of $R_{\alpha 1}(y)$ and $R_{\alpha 2}(y)$.  
Later we will also use the special notation $\text{KL}_1(y)$, $\text{KL}_2(y)$, $p_{\text{KL}1}$ and $p_{\text{KL}2}$ for $\lim_{\alpha\rightarrow 1} R_{\alpha 1}(y)$, 
$\lim_{\alpha\rightarrow 1}R_{\alpha 2}(y)$, $\lim_{\alpha \rightarrow 1}p_{\alpha 1}$ and $\lim_{\alpha\rightarrow 1} p_{\alpha 2}$.  
As mentioned earlier, the limit $\alpha\rightarrow 1$ in the R\'{e}nyi divergence corresponds to the Kullback-Leibler divergence.  

There are a number of ways that the basic approach above can be modified.  One possibility is to replace the posterior distribution $g(\theta_2|y_{\text{obs}})$ in
the refence distribution (\ref{refklcheck1}) with an appropriate partial posterior distribution 
\cite{bayarri+b00,bayarri+c07} $g(\theta_2|y_{\text{obs}}\backslash R_{\alpha 1}(y_{\text{obs}}))$ defined for data $y$ by 
$$g(\theta_2|y \backslash R_{\alpha 1}(y)) \propto g(\theta_2)\frac{p(y|\theta_2)}{p(R_{\alpha 1}(y)|\theta_2)}.$$
The partial posterior removes the information in $R_{\alpha 1}(y)$ about $\theta_2$ from the likelihood $p(y|\theta_2)$  
in calculating a reference posterior for $\theta_2$
for use in (\ref{refklcheck1}).  We would also use the partial posterior in taking the expectation in (\ref{Ralpha1}).  
To get some intuition, imagine receiving the information in $y$ in two pieces where we are told the value of $R_{\alpha 1}(y)$ first, followed
by the remainder;  if we applied Bayes' rule sequentially, first updating the prior $g(\theta_2)$ by $p(R_{\alpha 1}(y)|\theta_2)$, then the ``likelihood" term needed to update the posterior
given $R_{\alpha 1}(y)$ to the full posterior $g(\theta_2|y)$ would be $\frac{p(y|\theta_2)}{p(R_{\alpha 1}(y)|\theta_2)}$.  So the partial posterior just
updates the prior for $g(\theta_2)$ by this second likelihood term that represents the information in the data with that from $R_{\alpha 1}(y)$ removed.  
This somehow avoids an inappropriate double use of the data where the same information is being used to both construct a reference distribution and assess lack of fit.  
Use of the partial posterior distribution makes computation of (\ref{klcheck1}) more complicated, however.

There are some other ways that the basic hierarchically structured check can be modified in some problems with additional
structure.  In their discussion of checking hierarchical priors, \shortciteN{evans+m06} consider two situations.  
The first situation is where the likelihood is a function $\theta_1$ only, $p(y|\theta)=p(y|\theta_1)$.  
In this case, suppose that $T$ is a minimal sufficient statistic for $\theta_1$ in the model
$p(y|\theta_1)$ and that $V=V(T)$ is minimal sufficient for $\theta_2$ in the marginalized model $\int p(y|\theta_1)p(\theta_1|\theta_2)\;d\theta_1$.  
Writing $t_{\text{obs}}$ and $v_{\text{obs}}$ for the observed values of $T$ and $V$, 
they suggest further decomposing the term $p(t)$ in (\ref{decomposition}) as $p(v)p(t|v)$ where $p(v)$ denotes the prior predictive density for $V$ and
$p(t|v)$ denotes the prior predictive density for $T$ given $V=v$.  In this decomposition it is suggested that $p(t|v)$ should be used for checking 
$g(\theta_1|\theta_2)$, by comparing $p(t_{\text{obs}}|v_{\text{obs}})$ with $p(T|v_{\text{obs}})$ for draws of $T$ from $p(t|v_{obs})$, and then
if no conflict is found 
$p(v)$ should then be used for checking $g(\theta_2)$, by comparing $p(v_{obs})$ with $p(V)$ for $V\sim p(v)$.  
So checking $g(\theta_2)$ should be based on the prior predictive for $V$ and checking $g(\theta_1|\theta_2)$ should be based on 
a statistic that is a function of $T$ with reference distribution that of the conditional for $T|V=v_{obs}$ induced under the prior predictive for the data.
Looking at our hierarchically structured check, if there exists a minimal sufficient statistic $V$ for $\theta_2$, then we see in (\ref{klcheck2}) our
checking statistic $R_{\alpha 2}(y)$ is a function of that statistic and it will be invariant to what minimal sufficient statistic is chosen.  We are also using the prior
predictive for the reference distribution so our approach fits nicely with that of \shortciteN{evans+m06}.  
In the check (\ref{klcheck1}) we can see that the model checking statistic is a function of $T$ and invariant to the choice of $T$.  If we were to change
the reference distribution (\ref{refklcheck1}) to that of $T|V=v_{obs}$ then (\ref{klcheck1}) would also fit naturally with the approach
of \shortciteN{evans+m06}.  
However, sometimes suitable non-trivial sufficient statistics are not available and the conditional prior predictive of $T$ given $V=v_{obs}$ might be difficult to work with.  
Our general approach of using the posterior distribution of $\theta_2$ given $v_{\text{obs}}$ to integrate out $\theta_2$
comes close to achieving the ideal considered in \shortciteN{evans+m06} when there are sufficient
statistics at different levels of the model.  
A final observation is that we could consider a cross-validatory version of the check if interest
centred on a certain observation specific parameter within the vector $\theta_1$.  This approach is considered further
in a later example.

The other situation considered in \shortciteN{evans+m06} for checking hierarchical priors is the case where $p(y|\theta)$ can depend on both 
$\theta_1$ and $\theta_2$.  Here they suppose there is some minimal sufficient $T$ and a maximal ancillary statistic $U(T)$ for
$\theta$, and a maximal ancillary statistic $V$ for $\theta_1$ (ancillary for $\theta_1$ means that the sampling distribution of $V$ given $\theta$ depends only on $\theta_2$). 
Conditioning on ancillaries is relevant since we don't want assessment of prior-data conflict to depend on variation
in the data that does not depend on the parameter.   
They suggest in (\ref{decomposition}) decomposing $p(t)$ as $p(u)p(v|u)p(t|v,u)$ and using the second term $p(v|u)$ (the conditional distribution of 
$V$ given $U$ induced under the prior predictive for the data) to check $g(\theta_2)$, with the third term $p(t|v,u)$ (the conditional distribution of 
$T$ given $V$ and $U$ under the prior predictive for the data) used to check $g(\theta_1|\theta_2)$.  Again we can modify our suggested approach
where this additional structure is available.  If we change $g(\theta_2|y)$ to $g(\theta_2|v)$ in the definition of $R_{\alpha 2}(y)$, 
then we are checking $g(\theta_2)$ using a discrepancy which is a function of $V$.  If no maximal ancillary for $\theta$ were available, the suggestion
of \shortciteN{evans+m06} would use the prior predictive for $V$ for the reference distribution.  Because $V$ is ancillary for $\theta_1$ 
the check does not depend in any way on $g(\theta_1|\theta_2)$, which is desirable because we would like to check for conflict with $\theta_2$ separately
from checking for any conflict with $g(\theta_1|\theta_2)$.  For the check (\ref{klcheck1}) our discrepancy is a function of $T$ as \shortciteN{evans+m06} would recommend, 
and if the reference predictive distribution were changed to be that of $T$ given $U$ and $V$ we could use this approach to check for conflict with
$g(\theta_1|\theta_2)$.  However, in complex situations identifying suitable maximal ancillary statistics may not be possible.   Nevertheless consideration of 
problems like this provides some guidance as an ideal.

\subsection{Other suggestions for prior-data conflict checking}

Now that we have given the basic idea of our method we discuss its connections with other suggestions in the literature.
Perhaps the approach to prior-data conflict detection most closely related to the one developed here has been suggested by \shortciteN{bousquet08}.  Similar to 
us, \shortciteN{bousquet08}
considers a test statistic based on  prior to posterior (Kullback-Leibler) divergences, but uses the ratio of two such divergences.  Briefly, a non-informative prior 
is defined and then a reference posterior distribution for this non-informative prior is constructed.  Then, the prior to reference posterior divergence for the prior to be
examined is computed and divided by the prior to reference posterior divergence for the non-informative prior.  
When the non-informative prior is improper, some modification of the basic procedure is suggested, and extensions to hierarchical settings are also discussed.  
The approach we consider here has similar intuitive roots but is simpler
to implement because it does not require the existence of a non-informative prior.   We consider the prior
to posterior divergence for the prior under examination, a measure of how much beliefs have changed from prior to posterior,  
and compare the observed value of this statistic to its distribution under the prior predictive for the data.  
There is hence no need to define a non-informative prior, although as mentioned earlier there are interesting asymptotic connections between the checks
we suggest and Jeffreys' and reference non-informative priors.  
This will be discussed further in Section 4.  
Our focus here is not on deriving non-informative prior choices, however, but on
detecting conflict for a given proper prior.  

A quite general and practically implementable suggestion for measuring prior-data conflict has been given recently by \shortciteN{presanis+osd13}.  
Their approach generalizes earlier work by \shortciteN{marshall+s07} and also relates closely to some previous suggestions by 
\shortciteN{gasemyr+n09} and \shortciteN{dahl+gn07}.  They give a general conflict diagnostic that can be applied to a node or group of nodes of a model specified as
a directed acyclic graph (DAG).  The conflict diagnostic is based on formulating two distributions representing independent sources of information about the separator
node or nodes which are then compared.  Again, in general, there is a need in this approach to specify non-informative priors for the purpose
of formulating distributions representing independent sources of information.  \shortciteN{ohagan03} is an earlier suggestion for
examining conflict at any node of a DAG that was inspirational for much later work in the area, although the specific procedure 
suggested has been found to suffer from conservatism in some cases.  \shortciteN{scheel+gr11} consider a graphical approach to examining conflict
where the location of a marginal posterior distribution with respect to a local prior and lifted likelihood is examined, where the local prior and lifted likelihood are
representing different sources of information coming from above and below the node in a chain graph model.  
\shortciteN{reimherr+mn14} examine prior-data conflict by considering the difference
in information in a likelihood function that is needed to obtain the same posterior uncertainty for a given proper prior compared to a baseline prior.  Again, 
some definition of a non-informative prior for the baseline is needed for this approach to be implemented.  
Finally the model checking approach considered in \shortciteN{dey+gsv98} can also be used for checking for prior-data conflict.  There is some similarity
with our approach in that they use quantities associated with the posterior itself in the test.  Specifically they consider Monte Carlo tests based on vectors
of posterior quantiles and the prior predictive with a Euclidean distance measure used to measure similarity between the vectors of
quantiles.  

\section{First examples}

To begin exploring the properties of the conflict check (\ref{klcheck}), we consider a series of simple examples where calculations can be done analytically.
These examples were also given in \shortciteN{evans+m06}, and we compare with their check (\ref{emcheck}) in each case.

\begin{exmp}
{\it Normal location model}.  

\noindent
Suppose $y_1,\dots,y_n\sim N(\mu,\sigma^2)$ where $\mu$ is an unknown mean and $\sigma^2>0$ is a known variance.  
In this normal location model the sample mean is sufficient for $\mu$ and normally distributed
so without loss of generality we may consider $n=1$ and write the observed data point as $y_{\text{obs}}$.  
The prior density $g(\mu)$ for $\mu$ will be assumed normal, $N(\mu_0,\sigma_0^2)$ where $\mu_0$ and $\sigma_0^2$ are known.  

To implement the conflict check of \shortciteN{evans+m06} we need $p(y)$ which is normal, $N(\mu_0,\sigma^2+\sigma_0^2)$ (the sufficient statistic
in this case of a single observation is just $y$). 
Here and in later examples we use the notation $A(y)\doteq B(y)$ to mean that $A(y)$ and $B(y)$ are related (as a function of $y$) by a monotone transformation.
When conducting a Bayesian model check with discrepancies $D_1(y)$ and $D_2(y)$ 
then they will result in the same predictive $p$-values if $D_1(y)\doteq D_2(y)$ (although care must be taken to compute the appropriate left or right tail area, since in
our definition of the $\doteq$ notation the relationship between $A(y)$ and $B(y)$ can be either monotone increasing or decreasing).  
Now we can write 
$\log p(y)\doteq (y-\mu_0)^2$ and  
we see that the check of \shortciteN{evans+m06} compares $(y_{\text{obs}}-\mu_0)^2$ to the distribution of $(Y-\mu_0)^2$ for $Y\sim p(y)$.  
Following the similar example of \shortciteN{evans+m06}, p. 897, the $p$-value is 
$$p_{\text{EM}}=2\left(1-\Phi\left(\frac{|y_{\text{obs}}-\mu_0|}{\sqrt{\sigma^2+\sigma_0^2}}\right)\right).$$
Next, consider the prior-data conflict check based on the R\'{e}nyi divergence statistic.  The posterior density for $\mu$ is $N(\tau^2 \gamma, \tau^2)$ where
$\tau^2=(1/\sigma_0^2+1/\sigma^2)^{-1}$ and $\gamma=(\mu_0/\sigma_0^2+y/\sigma^2)$
and the prior to posterior R\'{e}nyi divergence of order $\alpha$ is (using, for example, the formula in \shortciteN{gil+al13}),
$$R_\alpha(y)=\log \frac{\sigma_0}{\tau}+\frac{1}{2(\alpha-1)}\log \frac{\sigma_0^2}{\sigma_\alpha^2}+\frac{1}{2}\frac{\alpha (\tau^2\gamma-\mu_0)^2}{\sigma_\alpha^2},$$
where $\sigma_\alpha^2=\alpha\sigma_0^2+(1-\alpha)\tau^2$.  Here only $\gamma$ depends on $y$, so that 
$$R_\alpha(y)\doteq (\tau^2\gamma-\mu_0)^2\doteq (\gamma-\mu_0/\tau^2)^2=(y-\mu_0)^2/\sigma^2 \doteq (y-\mu_0)^2$$
and the divergence based check is equivalent to the check of \shortciteN{evans+m06} in this example for every value of $\alpha$.
\end{exmp}

\begin{exmp}
{\it Binomial model}

\noindent
Suppose that $y\sim \text{Binomial}(n,\theta)$ and write $y_{\text{obs}}$ for the observed value.  The prior density $g(\theta)$ of $\theta$ is Beta$(a,b)$, which for data $y$ results in the posterior density $g(\theta|y)$ being Beta$(a+y,b+n-y)$.  Using the expression for the R\'{e}nyi divergence between two beta distributions \cite{gil+al13}
\begin{align}
 R_\alpha(y) = & \log \frac{B(a,b)}{B(a+y,b+n-y)}+\frac{1}{\alpha-1}\log \frac{B\Bigl(a+\alpha y, b+\alpha (n-y)\Bigr)}{B(a+y,b+n-y)} \nonumber \\
 = & T_1+T_2 \label{ralphay}
\end{align}
where $B(\cdot,\cdot)$ denotes the beta function.  
Now consider the check of \shortciteN{evans+m06}.  
$y$ is minimal sufficient and
the prior predictive for $y$ is beta-binomial, 
\begin{align*}
 p(y) = & \binom{n}{y}\frac{B(a+y,b+n-y)}{B(a,b)}, \;\;\;\;\;y=0,\dots,n.
\end{align*}
Hence a suitable discrepancy for the check of \shortciteN{evans+m06}, which we denote by $\text{EM}(y)$, is
\begin{align}
  \text{EM}(y) & = \log p(y) \nonumber \\
  & = \log \binom{n}{y}+\log \frac{B(a+y,b+n-y)}{B(a,b)} \nonumber \\
   & \doteq \log\Gamma(a+y)+\log\Gamma(b+n-y)-\log\Gamma(y+1)-\log\Gamma(n-y+1).  \label{emstat}
\end{align}
The check of \shortciteN{evans+m06} and the divergence based check are not equivalent in this example.  However, they can be related to each other when $y$ and $n-y$ are both large.  
Using Stirling's approximation for the beta function 
$$B(x,z)\approx \sqrt{2\pi}\frac{x^{x-\frac{1}{2}}z^{z-\frac{1}{2}}}{(x+z)^{x+z-\frac{1}{2}}},$$
for $x$ and $z$ large, we obtain
\begin{align}
  T_1 \doteq & \log B(a,b)-(a+b+n)\hat{\theta}_n\log \hat{\theta}_n+\frac{1}{2}\log \hat{\theta}_n \nonumber \\
  & \;\;\;\;\;-(a+b+n)(1-\hat{\theta}_n)\log(1-\hat{\theta}_n)+\frac{1}{2}\log (1-\hat{\theta}_n)+O\left(\frac{1}{n}\right),  \label{term1}
\end{align}
where some constants not depending on $y$ have been ignored on the right hand side and 
$\hat{\theta}_n=(a+y)/(a+b+n)$ is the posterior mean of $\theta$. Another application of Stirling's approximation to to $T_2$ in (\ref{ralphay}) gives
\begin{align*}
 T_2 = & \frac{1}{\alpha-1}\log \frac{B\Bigl(a+\alpha y, b+\alpha (n-y)\Bigr)}{B(a+y,b+n-y)} \\
  = & \frac{1}{\alpha-1} \left\{\left(a+b+\alpha n\right)\tilde{\theta}_n \log\tilde{\theta}_n+\left(a+b+\alpha n\right)(1-\tilde{\theta}_n)\log (1-\tilde{\theta}_n) \right.  \\
     & \left. -\left(a+b+n\right)\hat{\theta}_n \log \hat{\theta}_n-\left(a+b+n\right)(1-\hat{\theta}_n)\log(1-\hat{\theta}_n)\right\}+O\left(\frac{1}{n}\right),
\end{align*}
where $\tilde{\theta}_n=(a+\alpha y)/\Bigl(b+n+\alpha n\Bigr)$.  Making the Taylor series approximations
\begin{align*}
 \tilde{\theta}_n\log \tilde{\theta}_n = & \hat{\theta}_n\log \hat{\theta}_n+(\tilde{\theta}_n-\hat{\theta}_n)(1+\log \hat{\theta}_n)+O\left(\frac{1}{n^2}\right), \\
(1-\tilde{\theta}_n)\log (1-\tilde{\theta}_n) = & (1-\hat{\theta}_n)\log(1-\hat{\theta}_n)-(\tilde{\theta}_n-\hat{\theta}_n)\Bigl(1+\log(1-\hat{\theta}_n)\Bigr)+O\left(\frac{1}{n^2}\right) 
\end{align*}
and also observing that $n(\tilde{\theta}_n-\hat{\theta}_n)=\frac{\alpha-1}{\alpha}\left\{(a+b)\hat{\theta}_n-a\right\}+O\left(\frac{1}{n}\right)$ gives
\begin{align}
  T_2 = & n\hat{\theta_n}\log\hat{\theta}_n+n(1-\hat{\theta}_n)\log(1-\hat{\theta}_n)+\Bigl((a+b)\hat{\theta}_n-a\Bigr)\log\hat{\theta}_n \nonumber \\
       & \;\;\;\;\;+\Bigl((a+b)\hat{\theta}_n-b\Bigr)\log(1-\hat{\theta}_n)+O\left(\frac{1}{n}\right). \label{term2}
\end{align}
Combining (\ref{term1}) and (\ref{term2}) gives
\begin{align*}
R_\alpha(y) \doteq & \log B(a,b)-\frac{1}{2}\log \hat{\theta}_n-\frac{1}{2}\log(1-\hat{\theta}_n)-(a-1)\log\hat{\theta}_n-(b-1)\log(1-\hat{\theta}_n)+O(1/n) \\
 \doteq & -\log g(\hat{\theta}_n)+\frac{1}{2}\log |I(\hat{\theta}_n)|+O(1/n)
\end{align*}
where $I(\theta)=n/(\theta(1-\theta))$ is the Fisher information and $g(\hat{\theta}_n)$ is the prior density evaluated at $\hat{\theta}_n$.  
The posterior mean can be replaced by any other estimator differing from it by $O(1/n)$ such as the maximum likelihood estimator.  We explain in Section 4 
why the form of the result above is expected much more generally.  

Turning now to the check of \shortciteN{evans+m06}, 
appropriate Taylor expansions in (\ref{emstat}) gives
\begin{align*}
 \log \Gamma (a+y) & = \log \Gamma(y+1)+(a-1)\psi(a+y) \\
 & = \log\Gamma(y+1)+(a-1)\log(a+y)+O(1/n), \\
\log \Gamma(b+n-y) & = \log \Gamma(n-y+1)+(b-1)\psi(b+n-y) \\
 & = \log \Gamma(n-y+1)+(b-1)\log(b+n-y)+O(1/n)
\end{align*}
which gives
\begin{align*}
 \log p(y)\doteq & \log \Gamma(y+1)+(a-1)\log (a+y)+\log\Gamma(n-y+1)+(b-1)\log (b+n-y) \\
 & \;\;\; -\log \Gamma(y+1)-\log\Gamma(n-y+1)+O(1/n) \\
 \doteq & (a-1)\log (a+y)+(b-1)\log (b+n-y)+O(1/n) \\
 \doteq & \log g(\hat{\theta}_n)+O(1/n),
\end{align*}
where as before $\hat{\theta}_n$ is the posterior mean for $\theta$.  
A general result about the check of \shortciteN{evans+m06} explaining the limiting form of the check above
is given in \shortciteN{evans+j11b}.
So the two checks differ asymptotically according to the presence of the term $-0.5 \log I(\hat{\theta}_n(y))$.  See the next Section for further discussion.

It is helpful to consider finite sample behaviour in some particular cases.  We see that for $R_\alpha(y)$ if we consider $\alpha\rightarrow\infty$, we obtain
\begin{align*}
\text{MR}(y) = & \log \frac{B(a,b)}{B(a+y,b+n-y)}+\frac{y}{n}\log \frac{y}{n}+\left(1-\frac{y}{n}\right)\log (n-y).
\end{align*}
If $a=b=1$ so that the prior is uniform, we see that
\begin{align*}
p_{\text{MR}} = & \frac{\# \left\{ y: \binom{n}{y}\left(\frac{y}{n}\right)^{y}\left(1-\frac{y}{n}\right)^{n-y}\geq \binom{n}{y_{\text{obs}}}\left(\frac{y_{\text{obs}}}{n}\right)^{y_{\text{obs}}} \left(1-\frac{y_{\text{obs}}}{n}\right)^{n-y_{\text{obs}}}\right\}}{n+1}
\end{align*}
and plotting $\binom{n}{y}\left(\frac{y}{n}\right)^{y}\left(1-\frac{y}{n}\right)^{n-y}$ reveals that it is symmetric with an antimode at $n/2$ when $n$ is
even and at $\{(n+1)/2,1+(n+1)/2\}$ when $n$ is odd.  So prior-data conflict is detected whenever $y_{\text{obs}}$ is near $0$ or $n$.  This does seem strange
when the prior is uniform but is perhaps not surprising given the asymptotic connection between our checks and the Jeffreys' prior, which is also not uniform in
this example.  
On the other hand note that, letting $p(m)$ denote the prior predictive density of $\text{MR}(y)$, then $p(m)=2/(n+1)$ when $n$ is even for all
$m$ except when $m$ is the antimode and
when $n$ is odd then $p(m)=1/(n+1)$ for all $m$.  So if we were to check the prior using $p(m)$ as the discrepancy rather than $\text{MR}(y)$ the $p$-value would
never be small and any conflict would be avoided.  

\end{exmp}

\begin{exmp}
{\it Normal location-scale model, hierarchically structured check}

\noindent
Extending our previous location normal example, suppose $y_1,\dots,y_n$ are independent $N(\mu,\sigma^2)$ where now both $\mu$ and $\sigma^2$ are unknown.  
Write $y=(y_1,\dots,y_n)$.  We consider a normal inverse gamma prior for $\theta=(\mu,\sigma^2)$, 
$\text{NIG}(\mu_0,\lambda_0,a,b)$ say, having density of the form 
$$g(\theta)=\frac{\sqrt{\lambda_0}}{\sigma\sqrt{2\pi}}\frac{b^a}{\Gamma(a)}\left(\frac{1}{\sigma^2}\right)^{a+1}\exp\left(-\frac{2b+\lambda_0(\mu-\mu_0)^2}{2\sigma^2}\right).$$
This prior is equivalent to $g(\theta)=g(\theta_2)g(\theta_1|\theta_2)=g(\sigma^2)g(\mu|\sigma^2)$ with $g(\sigma^2)$ inverse gamma, $\text{IG}(a,b)$ and $g(\mu|\sigma^2)$ normal, 
$N(\mu_0,\sigma^2/\lambda_0)$.  
In this model a sufficient statistic is $T=(\bar{y},s^2)$ where $\bar{y}$ denotes the sample mean and $s^2$ the sample variance and we write $t_{\text{obs}}=(\bar{y}_{\text{obs}},s^2_{\text{obs}})$ for its observed value.  The normal inverse gamma
prior is conjugate, and the posterior is $\text{NIG}(\mu_0'(y),\lambda_0',a',b'(y))$ where $\mu_0'(y)=(n+\lambda_0)^{-1}(\mu_0\lambda_0+n\bar{y})$, 
$\lambda_0'=n+\lambda_0$, $a'=(a+n/2)$ and 
$b'=b'(y)=b+(n-1)s^2/2+n(\bar{y}-\mu_0)^2/(2(n/\lambda_0+1))$.  
It is natural to consider the hierarchical checks we discussed earlier for testing the two components of $g(\theta)$.  First, let's consider
the check for conflict with $g(\mu|\sigma^2)$.  Using the expression for the R\'{e}nyi divergence between normal densities we get
\begin{align*}
 R_\alpha(y,\sigma^2)=& \log \frac{\lambda_0'}{\lambda_0}+\frac{1}{2(\alpha-1)} \log \frac{{\lambda_0'}^2}{\lambda_0^2}+\frac{1}{2}\frac{\alpha(\mu_0'(y)-\mu_0)^2}{\sigma_\alpha^2}
\end{align*}
where $\sigma_\alpha^2=\alpha\sigma_0^2/\lambda_0+(1-\alpha)\sigma^2/{\lambda_0'}^2$ and we note that
$$R_{\alpha 1}(y)\doteq \Bigl(\mu_0'(y)-\mu_0\Bigr)^2\doteq (\bar{y}-\mu_0)^2.$$
Our suggested hierarchical check compares $R_{\alpha 1}(y_{\text{obs}})$ to a reference distribution based on 
$Y\sim m(y)=\int p(\sigma^2|y_{obs})\int p(y|\mu,\sigma^2)p(\mu|\sigma^2)\;d\mu\;d\sigma^2$
Noting that the distribution of $\bar{y}$ under $m(y)$ is
$t_{2a'}\left(\mu_0,\sqrt{\frac{b'(y_{\text{obs}})}{a'}\left(\frac{1}{\lambda_0}+\frac{1}{n}\right)}\right)$ we see that the divergence based check just computes whether
$$\frac{\bar{y}_{\text{obs}}-\mu_0}{\sigma^*}=\frac{\bar{y}_{\text{obs}}-\mu_0}{\sqrt{b'(y_{\text{obs}})/a'(1/\lambda_0+1/n)}}$$ 
is larger in magnitude than a $t_{2a'}(0,1)$ variate.
The hierarchical check of \shortciteN{evans+m06}, p. 909, on the other hand calculates the probability that $(\bar{y}_{\text{obs}}-\mu_0)/\tilde{\sigma}$ is
larger in magnitude than a $t_{2a'-1}(0,1)$ variate, where $\tilde{\sigma}^2=(1/\lambda_0(n/\lambda_0+1)(2b+(n-1)s_{\text{obs}}^2))/(n/\lambda_0(n+2a-1))$.  
Clearly these checks are very similar, since both $\sigma^*$ and $\tilde{\sigma}$ are approximately $s/\sqrt{\lambda_0}$ for large $n$ and there is only one degree of freedom difference in the reference $t$-distribution. We also note that in our check if we change the reference distribution to be that of $y$ given $s^2$ (noting that $s^2$ is ancillary for $\mu$ 
and following the discussion of Section 2.2) then our check would then coincide with that of \shortciteN{evans+m06}.

Consider next the check on $p(\sigma^2)$.  
For two inverse gamma distributions, $p_1(\sigma^2)$ and $p_2(\sigma_2)$, being $\text{IG}(a',b')$ and $\text{IG}(a,b)$ respectively, the R\'{e}nyi divergence between them is 
\begin{align*}
 & \log \left\{\frac{\Gamma(a) {b'}^{a'}}{\Gamma(a') b^a}\right\}+\frac{1}{\alpha-1}\log \left\{\frac{\Gamma(a_\alpha)}{\Gamma(a')}\frac{{b'}^{a'}}{{b_\alpha}^{a_\alpha}}\right\}.
\end{align*}
where $a_\alpha=a'\alpha+(1-\alpha) a$ and $b_\alpha=\alpha b'+(1-\alpha)b$.  
Since $a$, $b$ and $a'$ don't depend on the data, this gives
\begin{align*}
 R_{\alpha 2}(y) \doteq & a' \log b'+\frac{1}{\alpha-1}a'\log b'-\frac{1}{\alpha-1}a_\alpha \log b_\alpha.
\end{align*}
Using $\log b_\alpha=\log (\alpha b'+(1-\alpha)b)= \log \alpha b'+(1-\alpha)b/(\alpha b')+O(1/n)$ and collecting terms 
\begin{align*}
  R_{\alpha 2}(y) \doteq & \frac{a}{a'}\log b'+\frac{a_\alpha}{a' \alpha} \frac{b}{b'}+O\left(\frac{1}{n}\right)  \\
 \doteq & \log \frac{b'/a'}{b/a}+\frac{b/a}{b'/a'}+O\left(\frac{1}{n}\right).
\end{align*}
Note also that $s^2\approx b'/a'$ for large $n$, so that for large $n$ using $R_{\alpha 2}(y)$ as discrepancy is approximately the same
as using
\begin{align}
  & \log \frac{s^2}{b/a}+\frac{b/a}{s^2}.  \label{divergencecheck}
\end{align}

The check described in \shortciteN{evans+m06}, p. 910, compares $s^2/(b/a)$ to an $F_{n-1,2a}$ density.  
Plugging in $s^2/(b/a)$ to the expression for the log of the $F$ density, we have the statistic
\begin{align*}
 \text{EM}(y) \doteq & \frac{n-3}{2}\log \frac{s^2}{b/a}-\frac{n+2a-1}{2}\log \left(1+\frac{n-1}{2a}\frac{s^2}{b/a}\right),
\end{align*}
and then using the approximation $\log(1+x)\approx \log x + 1/x$ for large $x$ gives approximately
\begin{align*}
  \text{EM}(y) \doteq & \frac{n-3}{2} \log \frac{s^2}{b/a}-\frac{n+2a-1}{2} \log \left(\frac{s^2}{b/a}\right) -\frac{n+2a-1}{2}\frac{2a}{n-1}\frac{b/a}{s^2}+O\left(\frac{1}{n}\right) \\
  \doteq & -\frac{a-1}{2}\log \frac{s^2}{b/a} -\frac{n+2a-1}{n-1}\frac{b}{s^2}+O\left(\frac{1}{n}\right).
\end{align*}
So for large $n$, we have approximately
\begin{align*}
  \text{EM}(y) \doteq & \frac{a-1}{2a} \log \frac{s^2}{b/a}+\frac{b/a}{s^2},
\end{align*}
which, comparing with (\ref{divergencecheck}), clarifies the relationship to the divergence based check .  

\end{exmp}

\begin{exmp}
{\it A non-regular example}

\noindent
The following example is adapted from \shortciteN{jaynes76} and \shortciteN{li+sne16}.  Suppose we observe $y_1,\dots,y_n\sim f(y|\theta)$ where 
$f(y|\theta)=r\exp\Bigl(-r(y-\theta)\Bigr)I(y>\theta)$ where $r$ is a known parameter, $\theta>0$ is unknown and $I(\cdot)$ denotes the indicator function.  
We consider an exponential prior on $\theta$, $g(\theta)=\kappa \exp(-\kappa \theta)I(\theta>0)$.  Note that this is a non-regular example when inference
about $\theta$ is considered, due to the way that the support of the density for the data depends on $\theta$.  This means, for example, that the 
MLE as well as the posterior distribution are not asymptotically normal.  

The likelihood function is
\begin{align*}
 p(y|\theta)=c(y)\exp\Bigl(-nr(y_{\text{min}}-\theta)\Bigr)I(0<\theta<y_{min}),
\end{align*}
where $y_{\text{min}}$ denotes the minimum of $y_1,\dots,y_n$ and $c(y)=r^n\exp\Bigl(-nr(\bar{y}-y_{\text{min}})\Bigr)$ where $\bar{y}$ denotes the sample mean.  
A sufficient statistic is $y_{\text{min}}$, and its sampling distribution has density
\begin{align*}
 p(y_{\text{min}}|\theta) & = nr\exp\Bigl(-nr(y_{\text{min}}-\theta)\Bigr)I(0<\theta<y_{\text{min}}).
\end{align*}
The prior predictive of $y_{\text{min}}$ is
\begin{align}
  p(y_{\text{min}})= & nr\kappa \exp\left(-nry_{\text{min}}\right)\int_0^{y_{\text{min}}} \exp\left((nr-\kappa)\theta\right) d\theta \nonumber  \\
 = & \frac{nr\kappa}{nr-\kappa} \Bigl(\exp(-\kappa y_{\text{min}})-\exp(-nry_{\text{min}})\Bigr), \label{ppymin}
\end{align}
and this is the discrepancy for the test of \shortciteN{evans+m06}.  Consider now the 
statistic $R_\alpha(y)$.  We have $g(\theta|y)\propto \exp\Bigl((nr-\kappa)\theta\Bigr)I(0<\theta<y_{\text{min}})$ so that
\begin{align*}
  g(\theta|y) & = \frac{(nr-\kappa)}{\exp(t)-1}\exp\Bigl((nr-\kappa)\theta\Bigr)I(0<\theta<y_{\text{min}}),
\end{align*}
where $t=(nr-\kappa)y_{\text{min}}$.  Then
\begin{align*}
\int_0^{y_{\text{min}}} \left(\frac{g(\theta|y)}{g(\theta)}\right)^{\alpha-1}g(\theta|y)d\theta & = 
\frac{\kappa}{\alpha nr-\kappa}\left(\frac{(nr-\kappa)}{\kappa (\exp(t)-1)}\right)^{\alpha}\left[\exp\Bigl((\alpha nr-\kappa)y_{\text{min}}\Bigr)-1\right],
\end{align*}
and so
\begin{align*}
 R_\alpha(y) & = \frac{1}{\alpha-1}\log \frac{\kappa}{\alpha nr-\kappa}+\frac{\alpha}{\alpha-1}\log \left(\frac{(nr-\kappa)}{\kappa (\exp(t)-1)}\right) \\
 & \hspace{0.5in} +\frac{1}{\alpha-1}\log\Bigl(\exp\Bigl((\alpha nr-\kappa)y_{\text{min}}\Bigr)-1\Bigr).
\end{align*}
To simplify notation, we write $t=(nr-\kappa)y_{\text{min}}$ as $\kappa (\nu-1)y_{\text{min}}$, where $\nu=nr/\kappa$.  We write $t_{\text{obs}}$ for the observed value.  
Then the prior predictive for $t$ obtained
by a change of variables in (\ref{ppymin}) is
\begin{align*}
  p(t) = & \frac{\nu}{(\nu-1)^2}\left[\exp\left(-\frac{t}{\nu-1}\right)-\exp\left(-\frac{\nu t}{\nu-1}\right)\right]I(t>0).
\end{align*}
The $p$-value $p_\alpha$ is 
\begin{align*}
  p_\alpha=p_\alpha(y) = 1-\int_{t_1}^{t_2}\frac{\nu}{(\nu-1)^2} \left[\exp\left(-\frac{t}{\nu+1}\right)-\exp\left(-\frac{\nu t}{\nu-1}\right)\right] \;dt,
\end{align*}
where $t_1$ and $t_2$ are such that $R_\alpha(t_1)=R_\alpha(t_2)=R_\alpha(t_{\text{obs}})$ with $t_1<t_0<t_2$ and $t_0$ is 
the value of $t$ at which $R_\alpha(y)=R_\alpha(t)$ is minimal.  There is a single global minimum with $R_\alpha (t)$ decreasing for $t<t_0$ and increasing
for $t>t_0$.  Either $t_1$ or $t_2$ will be equal to $t_{\text{obs}}$.  We can easily see that if $t_{\text{obs}}=t_0$ then $p_\alpha=1$, and if 
$t_{\text{obs}}\rightarrow\infty$ then $p_\alpha\rightarrow 0$.  Figure \ref{fig1} considers the special case of the KL divergence and shows some plots
of how $p_{\text{KL}}$ varies with $t_{\text{obs}}$ for a few different values of $\nu=nr/\kappa$. 
\begin{figure}[htbp]
\centering
\begin{tabular}{cc}
\includegraphics[width=70mm]{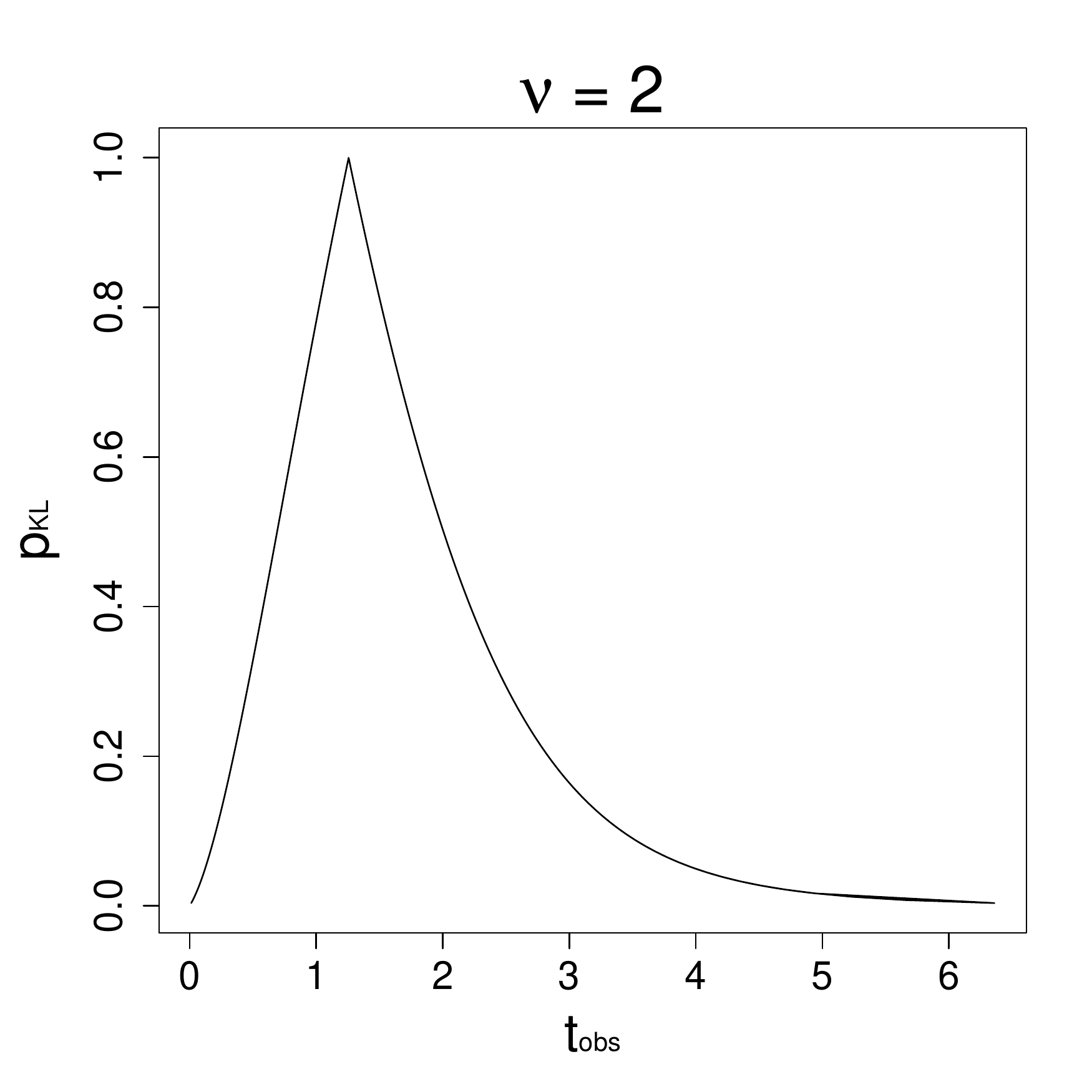} &
\includegraphics[width=70mm]{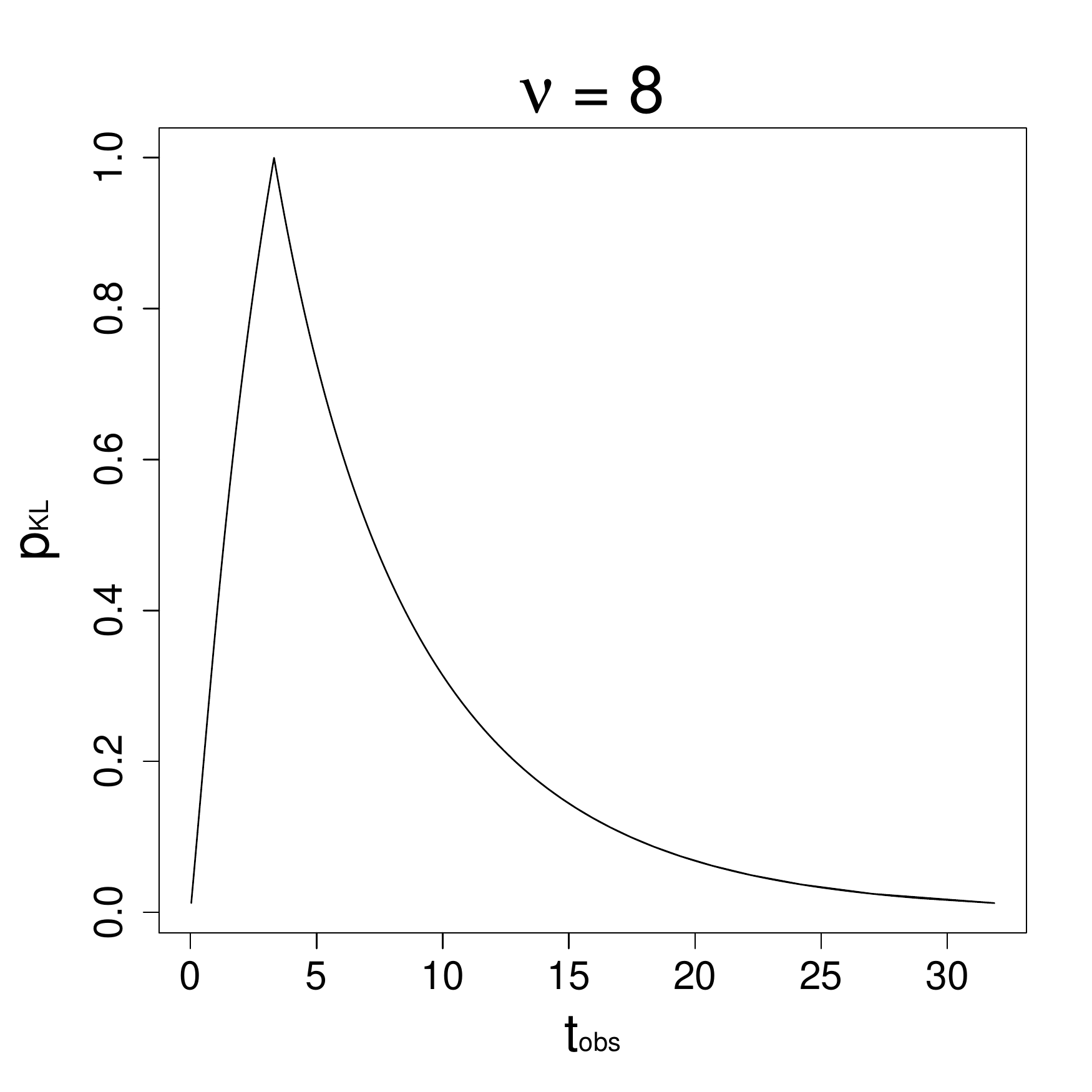}  \\
\multicolumn{2}{c}{\includegraphics[width=70mm]{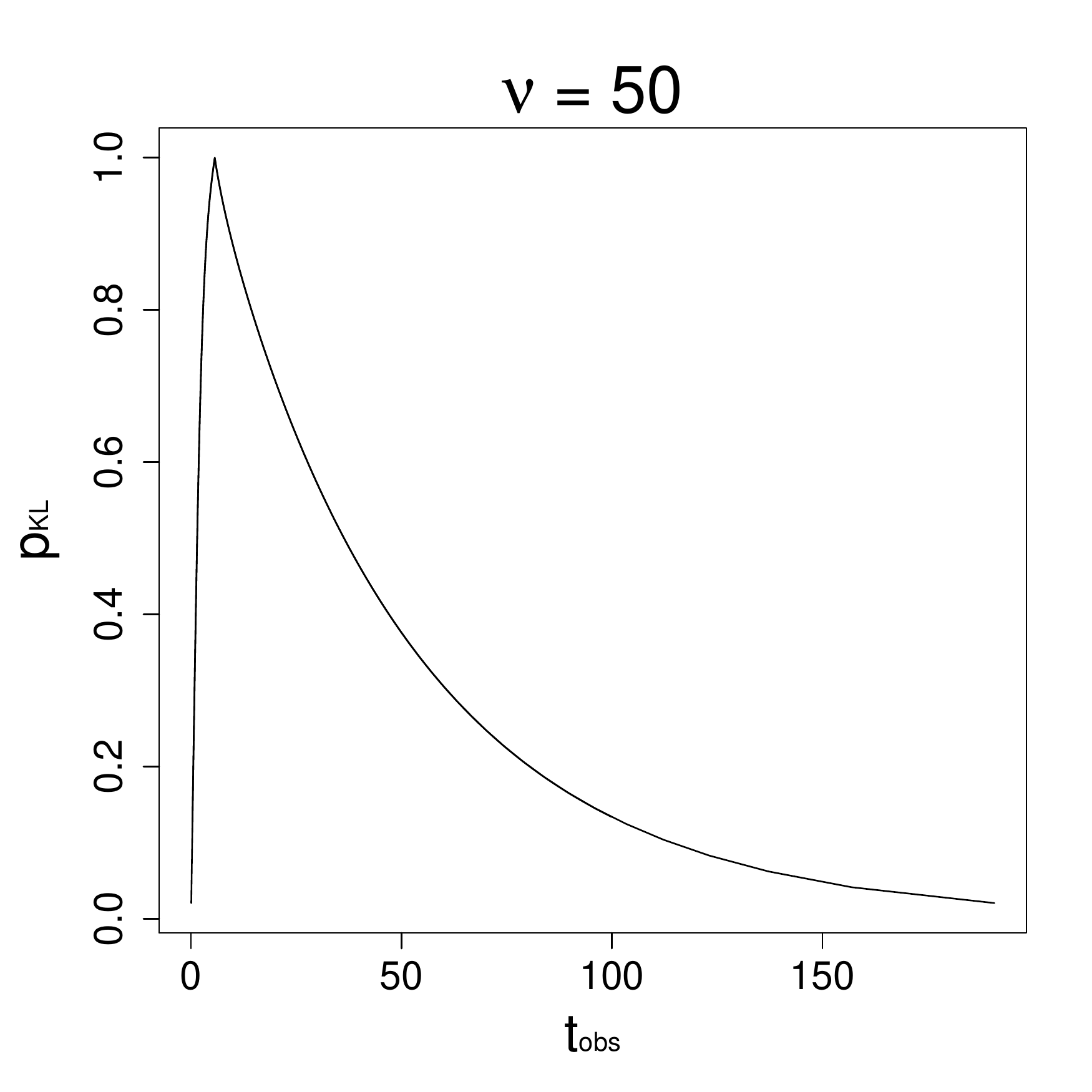} }
\end{tabular}
\caption{\label{fig1}Plots of $p_{\text{KL}}$ versus $t_{\text{obs}}$ for $\nu=2,8$ and $50$.}
\end{figure}

\end{exmp}

\section{Limiting behaviour of the checks}

We now give derivations of some of the limit results stated in Section 2.  
We will consider the special case of the Kullback-Leibler divergence first.  
Let $y_1,\dots,y_n$ be independent and identically distributed from $p(y|\theta)$ and denote
the true value of $\theta$ by $\theta^*$.  Write $n I(\theta)$ for the Fisher information and $n\hat{I}_n$ for the observed information.  Then under suitable regularity conditions
(see for example Theorem 1 of \shortciteN{ghosh11}, which summarizes the discussion in \shortciteN{ghosh+ds06};  see also \shortciteN{johnson70})
an asymptotic expansion of the posterior distribution gives
\begin{align*}
 \log g(\theta|y)+\frac{d}{2}\log\frac{2\pi}{n}-\frac{1}{2}\log |\hat{I}_n|+\frac{n(\theta-\hat{\theta}_n)^T \hat{I}_n(\theta-\hat{\theta}_n)}{2} & = O_p\left(\frac{1}{\sqrt{n}}\right)
\end{align*}
almost surely $P_{\theta^*}$.  Adding and subtracting $\log g(\theta)$ from the left hand side and taking expectation with respect to $g(\theta|y)$
gives
{\small
\begin{align*}
 \mbox{KL}(y)+\int \log g(\theta) g(\theta|y)+\frac{d}{2}\log \frac{2\pi}{n}-\frac{1}{2}\log |\hat{I}_n|+\int \frac{n(\theta-\hat{\theta}_n)^T \hat{I}_n (\theta-\hat{\theta}_n)}{2}g(\theta|y)d\theta & = O_p\left(\frac{1}{\sqrt{n}}\right)
\end{align*}}
and using the asymptotic normality of the posterior and noting that
$\hat{I}_n-I(\theta)$ converges to zero almost surely, and $\hat{\theta}_n$ converges to $\theta^*$ almost surely under the assumed regularity conditions, gives
\begin{align*}
  \mbox{KL}(y)+\log g(\theta^*)+\frac{d}{2}\log 2\pi e-\frac{1}{2}\log |I(\theta^*)| & = O_p\left( \frac{1}{\sqrt{n}} \right)
\end{align*}
Hence the $p$-value (\ref{klcheck}) converges as $n\rightarrow \infty$ to 
\begin{align*}
 P\left(\frac{1}{2}\log |I(\theta)|-\log g(\theta)\geq \frac{1}{2}\log |I(\theta^*)|-\log g(\theta^*)\right) & = P\Bigl(g(\theta^*)|I(\theta^*)|^{-1/2} \geq g(\theta)|I(\theta)|^{-1/2}\Bigr).
\end{align*}

Next, consider our hierarchical checks and the conflict $p$-values (\ref{klcheck1}) and (\ref{klcheck2}).  
The check (\ref{klcheck2}) is really just the same check as in the non-hierarchical case, but applied to the model and prior with $\theta_1$ integrated out so 
the limit is the same as in the non-hierarchical case with the Fisher information being that for the marginalized model $p(y|\theta_1)=\int p(y|\theta)p(\theta_2|\theta_1)$, 
provided that an appropriate asymptotic expansion of the marginal posterior is available.  
For the check (\ref{klcheck1}), the reference predictive distribution $m(y)$ converges to $p(y|\theta_2^*)=\int p(y|\theta)p(\theta_1|\theta_2^*)d\theta_1$ as 
$n\rightarrow\infty$ and in this model with $\theta_2=\theta_2^*$ fixed we will get the limiting $p$-value
\begin{align*}
 & P\Bigl(g(\theta_1^*|\theta_2^*)|I_{11}(\theta_1^*,\theta_2^*)|^{-1/2} \geq g(\theta_1|\theta_2^*) |I_{11}(\theta_1,\theta_2^*)|^{-1/2}\Bigr) 
\end{align*}
where $I_{11}(\theta)$ denotes the submatrix of $I(\theta)$ formed by the first $d_1$ rows and $d_1$ columns and $\theta_1\sim g(\theta_1|\theta_2^*)$.  
Just as the choice of $g(\theta)$ as the 
Jeffreys' prior results in a limiting $p$-value of $1$ in the non-hierarchical case, choosing $g(\theta)$ according to the two stage reference prior \cite{berger+bs09,ghosh11} results in 
both the limiting $p$-values corresponding to (\ref{klcheck1}) and (\ref{klcheck2}) being $1$.  This provides at least some heuristic reason why,
from the point of view of avoidance of conflict, a reference prior might be considered desirable.  It is not our intention here however to develop methodology for
default non-subjective prior choice or even to justify existing choices, but rather to develop methods for checking for conflict with given proper priors.  

Regarding the extension of the above ideas to the more general case of the R\'{e}nyi divergence, using a Laplace approximation to the integral
$$\int \left\{\frac{g(\theta|y)}{g(\theta)}\right\}^{\alpha-1} g(\theta|y)d\theta=\int g(\theta)^{-(\alpha-1)}g(\theta|y)^\alpha d\theta,$$
expanding about the mode $\hat{\theta}$ of $g(\theta|y)$ and replacing the Hessian of $\log g(\theta|y)$ at the mode with $n\hat{I}_n$, gives
\begin{align}
 & (2\pi)^{d/2}g(\hat{\theta}|y)^\alpha g(\hat{\theta})^{-(\alpha-1)} | \alpha n \hat{I}_n |^{-1/2},  \label{laplace}
\end{align}
and using the asymptotic normal approximation to $g(\theta|y)$, $N(\hat{\theta},n^{-1}\hat{I}_n^{-1})$, so that
\begin{align}
 g(\hat{\theta}|y) \approx & (2\pi)^{-d/2} | n \hat{I}_n|^{1/2},  \label{densmode}
\end{align}
and combining (\ref{laplace}) and (\ref{densmode}), gives
\begin{align*}
  R_\alpha(y) \approx & \frac{1}{\alpha-1}\left(-\frac{d}{2}\log 2\pi -\frac{\alpha d}{2}\log 2\pi+\frac{\alpha d}{2}\log n +\frac{\alpha}{2}\log |\hat{I}_n| \right. \\
 & \;\;\left.-(\alpha-1)\log g(\hat{\theta})-\frac{\alpha nd}{2}-\frac{1}{2}\log |\hat{I}_n|\right) \\
 \doteq & -\log g(\hat{\theta})+\frac{1}{2}\log |\hat{I}_n|,
\end{align*}
which converges to $-\log g(\theta)+\frac{1}{2}\log |I(\theta)|$ and hence we expect
a similar limit will hold for the $p$-value as for the Kullback-Leibler case, under suitable conditions.  

\section{More complex examples and variational Bayes approximations}

To calculate the check (\ref{klcheck}) or its hierarchical extensions may seem difficult.  Computation of $R_\alpha(y)$ involves an integral which is usually intractable, and an 
expensive Monte Carlo procedure may be needed to approximate it.  Furthermore, the integrand involves the posterior distribution.   
Even worse, as well as computing $R_\alpha(y_{\text{obs}})$, we need to compute a reference distribution for it, 
and this may involve calculating $R_\alpha(y^{(i)})$ for $y^{(i)}$, $i=1,\dots,m$, independently drawn from the prior predictive distribution.  So a straightforward Monte Carlo
computation of $p_{\alpha}$ may involve calculating $R_\alpha(y)$ for $m+1$ different datasets where $m$ might be large and with each of these calculations itself being expensive.
Here we suggest a way to make the computations easier using variational approximation methods.  
\shortciteN{tan+n14} also considered the use of variational approximations for computation of conflict diagnostics in hierarchical models and they show a relationship between
the diagnostics they consider and the mixed predictive checks of \shortciteN{marshall+s07}.  Their use of variational approximations for conflict detection
is very different to that considered here, however.  

In the variational approximation literature there are quite general
methods for learning approximations to the posterior that are in the exponential family (\shortciteNP{attias99,jordan+gjs99,winn+b05,rohde+w15}).  If the prior distribution for a certain block of parameters is also in the
same exponential family as its variational approximation, it is possible to compute the R\'{e}nyi divergence in closed form (\shortciteNP{liese+v87}).  
Furthermore, because variational approximations
are fast to compute, they are ideally suited to the repeated posterior computations
for samples under a reference predictive distribution that we need to compute $p_{\alpha}$.  

More generally there are also useful methods for learning approximations which are mixtures of Gaussians (\shortciteNP{salimans+k13,gershman+hb12}) 
and if the prior can also be approximated
by a mixture of Gaussians then useful closed from approximations to Kullback-Leibler divergences are available (\shortciteNP{hershey+o07}).  
We illustrate the use of variational methods for computing approximations of our conflict $p$-values 
in two examples.  In these examples we use the Kullback-Leibler divergence as the divergence measure.  In the first example we use a variational mixture
approximation, and in the second a Gaussian approximation in a hierarchically structured check for a logistic random effects model.

\begin{exmp}
{\it Beta-binomial example} \\
We consider the example in \shortciteN[Section 5.4]{albert2009bayesian}. This example estimates the rates of death from stomach cancer for males at risk aged $45-64$ for the $20$ largest cities in Missouri. The data set cancer mortality is available in the {\tt R} package {\tt LearnBayes} (\shortciteNP{albert2009bayesian}). It contains 20 observations denoted by $(n_i, y_i), i = 1, \ldots, 20$, where $n_i$ is the number of people at risk and $y_i$ is the number of deaths in the $i$th city.  An interesting model for these data is a beta-binomial with mean $\eta$ and precision $K$, where the probability function for the $i$th observation is
\begin{align*}
p(y_i|\eta, K) &= \binom{n_i}{y_i} \frac{B\Bigl(K\eta + y_i, K(1-\eta)+n_i-y_i\Bigr)}{B\Bigl(K\eta,K(1-\eta)\Bigr)}.
\end{align*}
\shortciteN{albert2009bayesian} considers the prior $g(\eta, K) \propto \frac{\displaystyle 1}{\displaystyle \eta (1 - \eta)}\frac{\displaystyle 1}{\displaystyle (1+K)^2}$ and
then reparametrizes to $\theta=(\theta_1,\theta_2)$ where
\begin{align*}
\theta_1 &= \mathrm{logit}(\eta) = \log \left(\frac{\eta}{1 - \eta}\right),\;\;\;\;\; \theta_2 = \log(K).
\end{align*}
We use this parametrization, but since Albert's prior on $(\eta,K)$ is improper we consider
a Gaussian prior for $\theta$, $g(\theta)=N(\mu_0,\Sigma_0)$, where $\mu_0$ is the mean and $\Sigma_0$ the
covariance matrix.  The posterior distribution $g(\theta|y)$ has a non-standard form, and we approximate it using a Gaussian mixture model (GMM).  
Variational computations are done using the algorithm in \shortciteN[Section 7.2]{salimans+k13} where the same dataset was also considered but with Albert's original prior.  
We consider a two component mixture approximation,
\begin{align*}
g(\theta|y) &\approx q(\theta) = \omega_1 q_1(\theta) + \omega_2 q_2(\theta),
\end{align*}
where $q(\theta)$ denotes the variational approximation, $\omega_1$ and $\omega_2$ are mixing weights with $\omega_1 + \omega_2 = 1$, and $q_1(\theta)$ and $q_2(\theta)$ are the normal mixture component densities with means and covariance matrices $\mu_1,\Sigma_1$ and $\mu_2,\Sigma_2$ respectively.   In our check, we replace
\begin{align*}
  \text{KL}(y) & = \int \log \frac{g(\theta|y)}{g(\theta)}g(\theta|y) d\theta
\end{align*}
with
\begin{align}
  \widetilde{\text{KL}}(y) & = \int \log \frac{q(\theta)}{g(\theta)}g(\theta|y)d\theta.  \label{statapprox1}
\end{align}
$\widetilde{\text{KL}}(y)$ replaces the true posterior $g(\theta|y)$ with its variational approximation.  Then we replace the exact  computation of (\ref{statapprox1}) with the closed form approximation of \shortciteN[Section 7]{hershey+o07}, which here takes the form
\begin{align*}
\omega_1 \cdot \log \frac{ \omega_1 + \omega_2 \cdot \exp\Bigl(-D(q_1||q_2)\Bigr)}{ \exp\Bigl(-D(q_1||g)\Bigr)} + \omega_2 \cdot \log \frac{ \omega_1 \cdot \exp\Bigl(-D(q_2||q_1)\Bigr) + \omega_2}{ \exp\Bigl(-D(q_2||g)\Bigr)},
\end{align*}
where $D(q_1||q_2)$, $D(q_1||g)$, $D(q_2||g)$ are the Kullback-Leibler divergences between $q_1$ and $q_2$, $q_1$ and $g$ and $q_2$ and $g$ respectively where $g$ is the prior.
There are closed form expressions for these Kullback-Leibler divergences since they are between pairs of multivariate Gaussian densities.  After application of the Hershey-Olsen bound, 
we have an approximating statistic $\text{KL}^*(y)$ to $\text{KL}(y)$.  
Then we can approximate $p_{\text{KL}}$ by simulating datasets $y^{(i)}$, $i=1,\dots,M$ under the prior predictive, computing $\text{KL}^*(y^{(i)})$ and $\text{KL}^*(y_{\text{obs}})$ and then 
$$p_{\text{KL}}\approx \frac{1}{M}\sum_{i=1}^M I\Bigl(\text{KL}^*(y^{(i)})\geq \text{KL}^*(y_{\text{obs}})\Bigr).$$
For illustration, consider three different normal priors, all with prior covariance matrix $\Sigma_0$ diagonal with diagonal entries $0.25$, 
but with prior means representing a lack of conflict, moderate conflict and a clear conflict ($\mu_0 = (-7.1, 7.9)$, $\mu_0=(-7.4, 7.9)$ and $\mu_0 = (-7.7, 7.9)$
respectively). Figure \ref{fig:lkhdpriorposterior} shows for the three cases contour plots of the prior and likelihood (left column) and the true posterior together with its
two component variational posterior approximation computed using the algorithm of \shortciteN{salimans+k13}. The three rows from top to bottom show the cases
of lack of conflict, moderate conflict and a clear conflict.  The $p$-values approximated by the variational method and Hershey-Olsen bound with $M=1000$ 
are $0.58$, $0.25$ and $0.03$ for the
three cases.  We can see that the variational posterior approximation is excellent even with just two mixture components and the $p$-values behave as we would expect.  \\
\begin{figure}[htp]
\begin{center}
\begin{tabular}{cc}
\includegraphics[height=66mm,width=63mm]{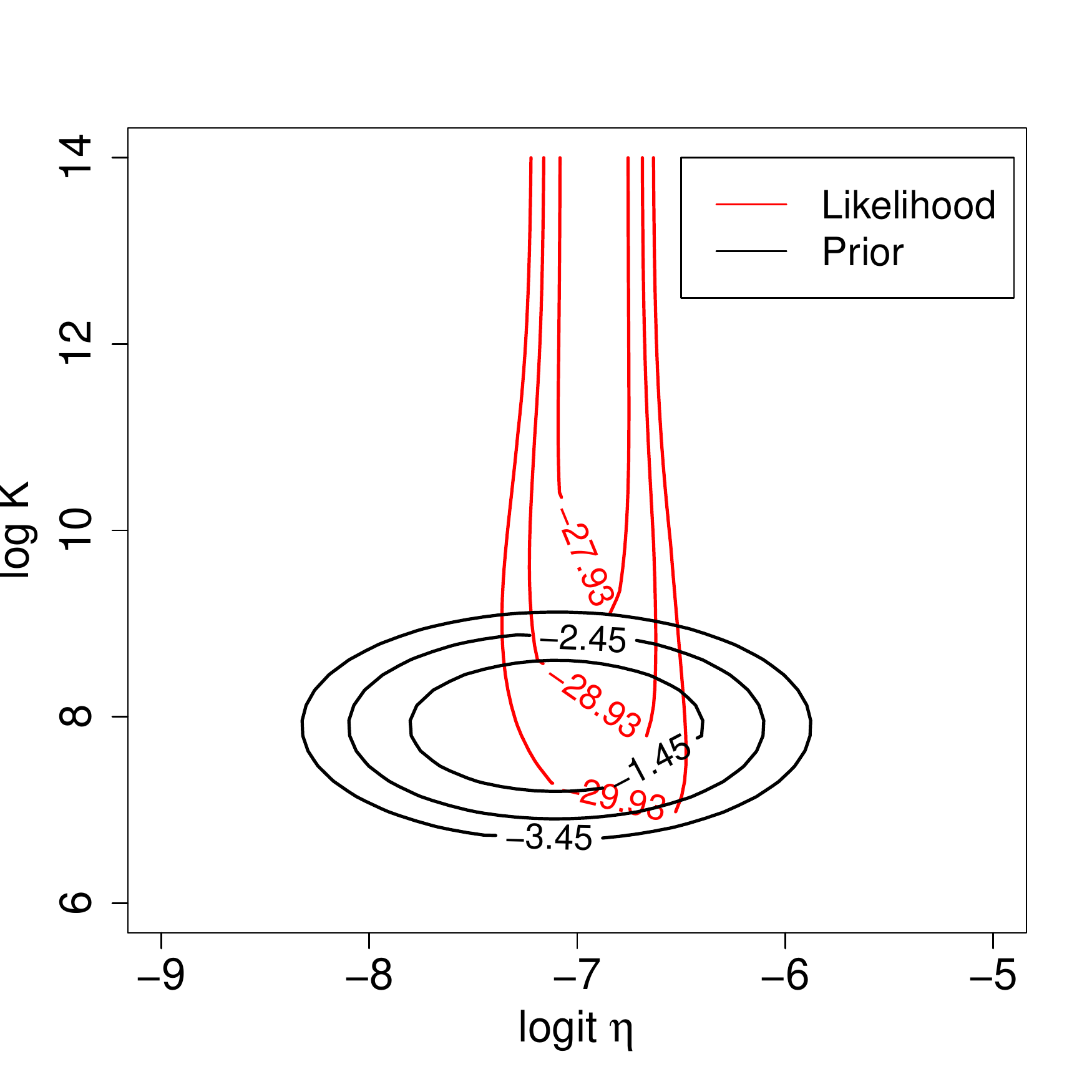}  & 
\includegraphics[height=66mm,width=63mm]{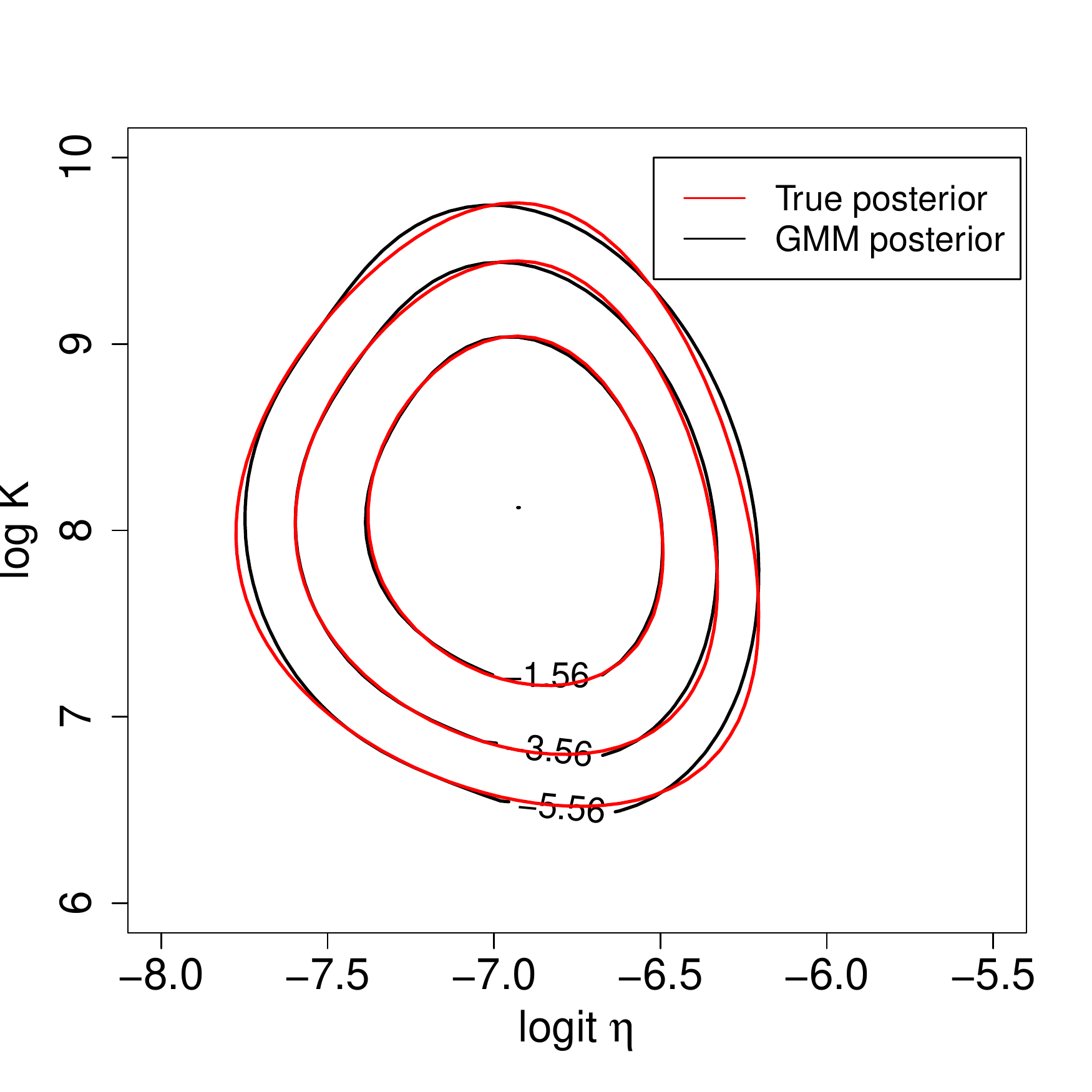} \\
\includegraphics[height=66mm,width=63mm]{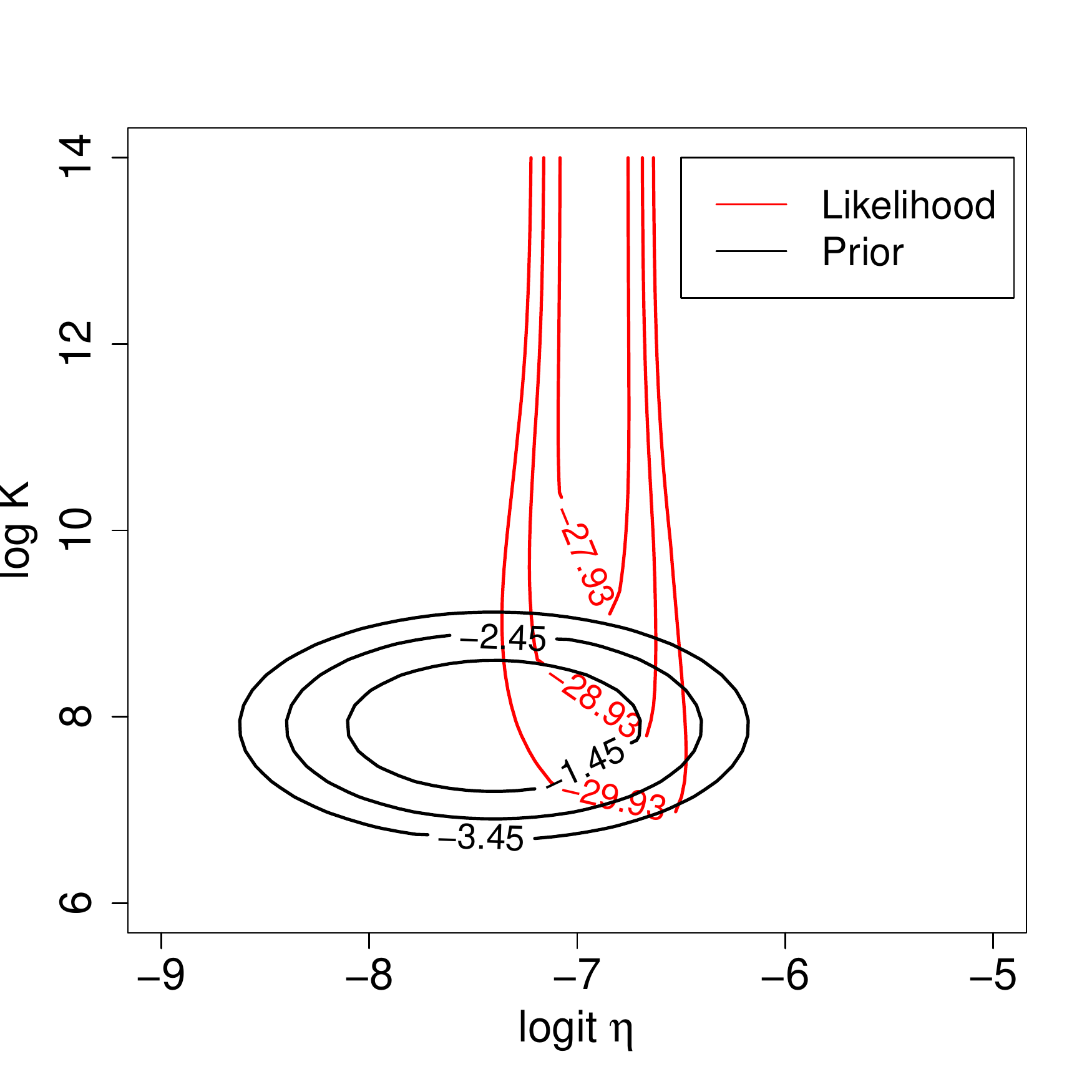}  & 
\includegraphics[height=66mm,width=63mm]{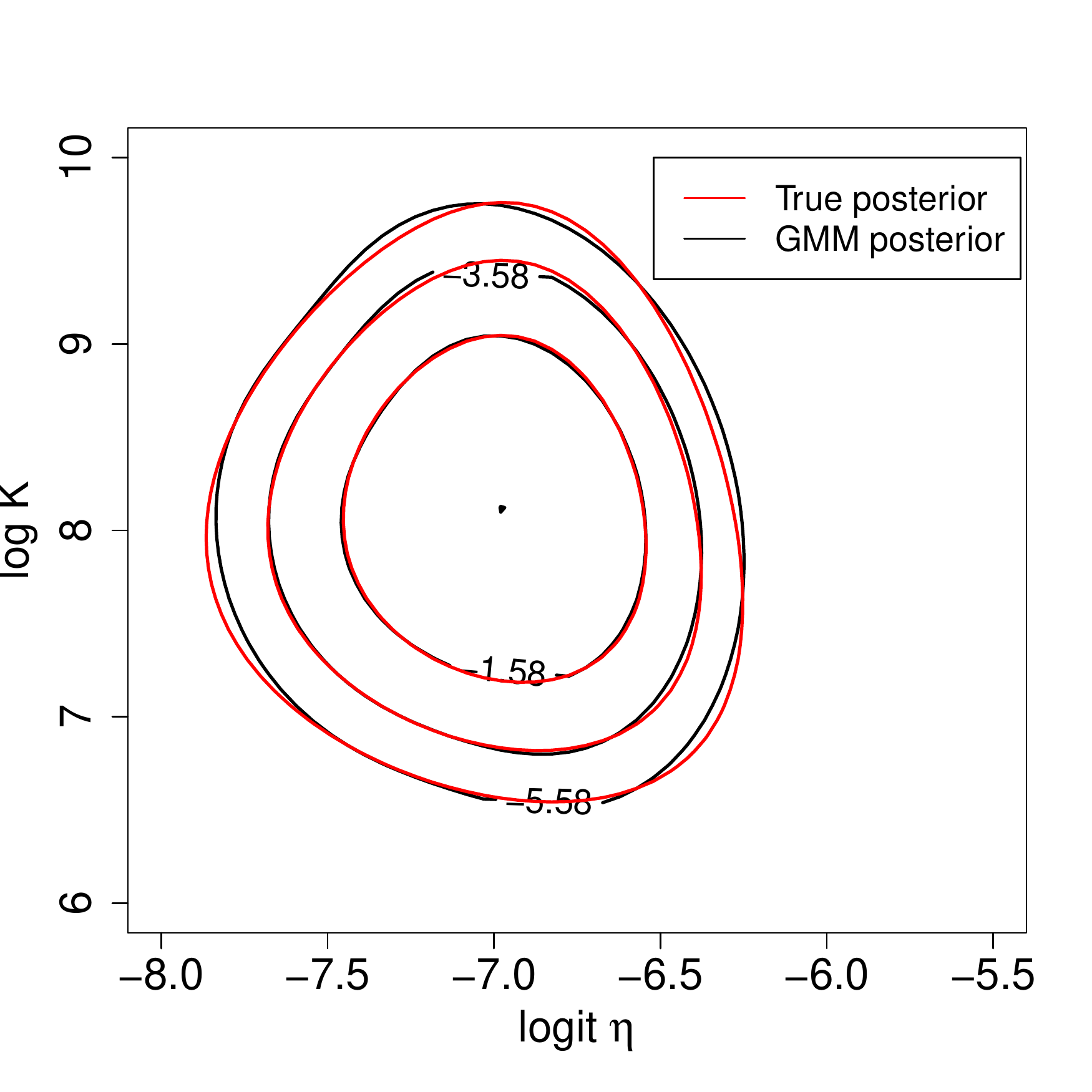} \\
\includegraphics[height=66mm,width=63mm]{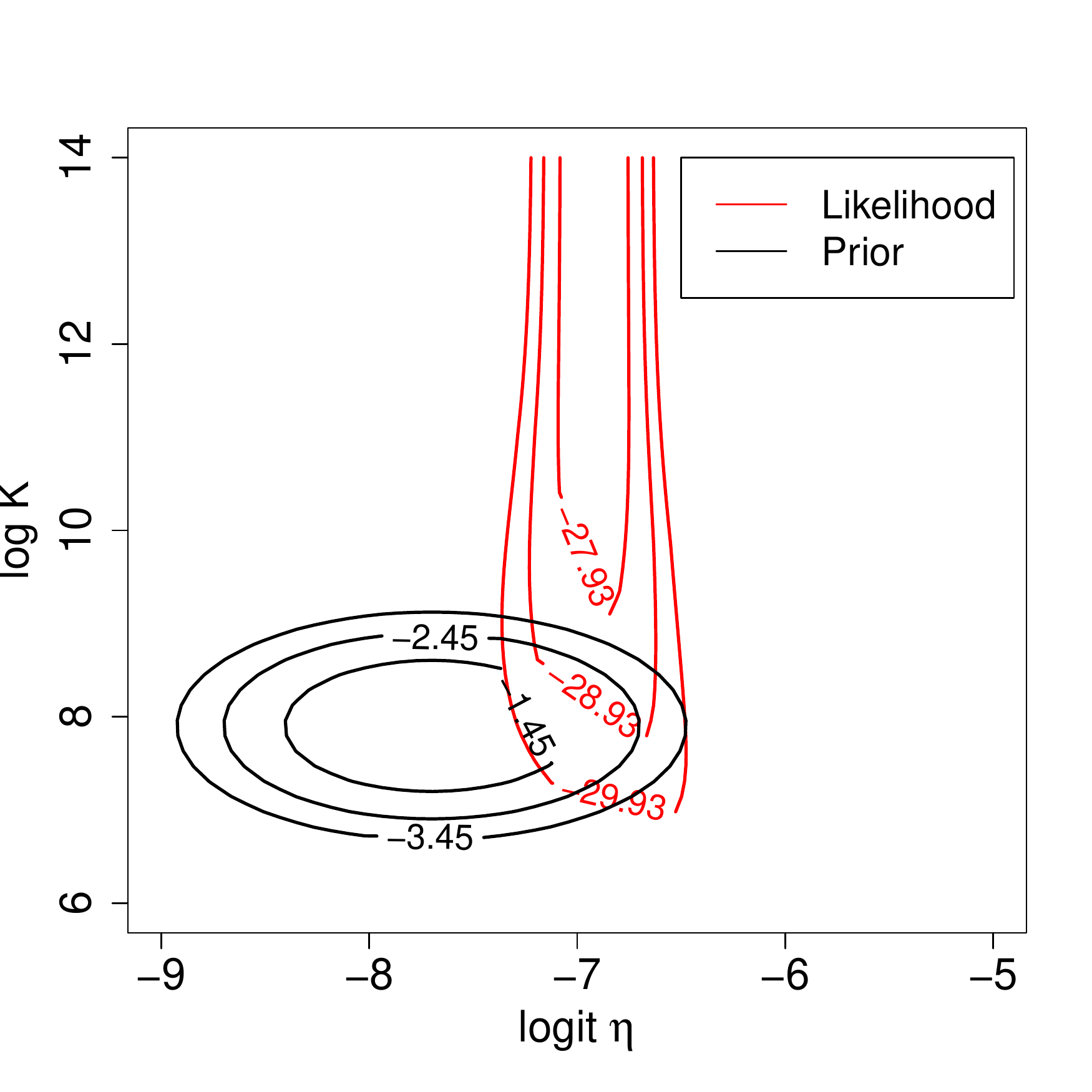}  & 
\includegraphics[height=66mm,width=63mm]{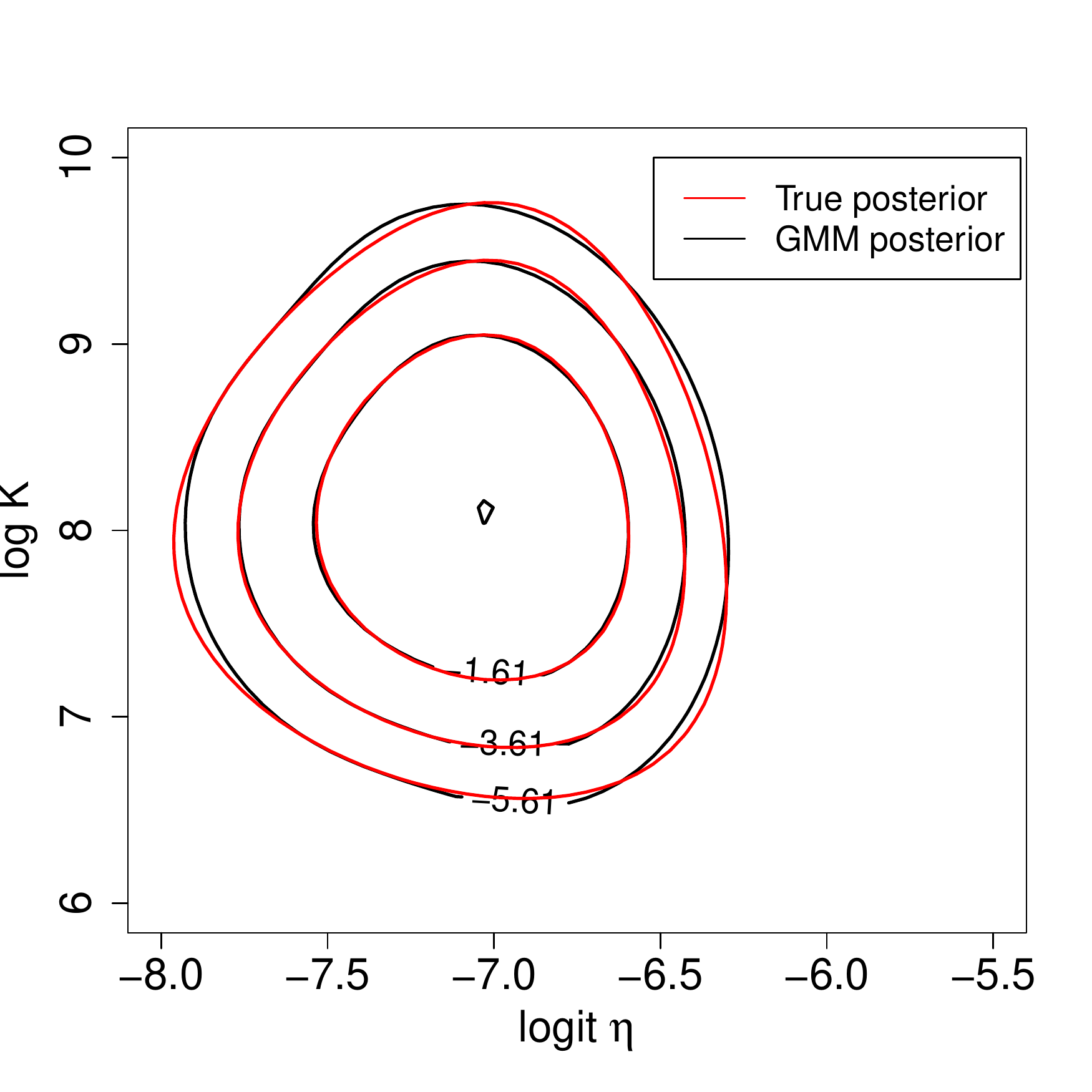} \\
\end{tabular}
\end{center}
\caption{\label{fig:lkhdpriorposterior} 
Contour plots of log-likelihood and prior (left) and true posterior together with Gaussian mixture approximation (right) for priors centered at $(-7.1, 7.9), (-7.4, 7.9)$ and $(-7.7, 7.9)$ (from top to bottom).
}
\end{figure}
\end{exmp}

\begin{exmp}
{\it Bristol Royal Infirmary Inquiry data} \\
We illustrate the computation of our conflict checks in a hierarchical setting using a logistic random effects model.  Here the data are part of that
presented to a public enquiry into excess mortality at the Bristol Royal Infirmary in complex paediatric surgeries prior to 1995.   The data are given in 
\shortciteN[Table 1]{marshall+s07} and a comprehensive discussion is given in \shortciteN{spiegelhalter+abem02}.  
The data consists of pairs $(y_i,n_i)$, $i=1,\dots,12$ where $i$ indexes different hospitals, $y_i$ is the number of deaths in hospital $i$ and $n_i$ is the number
of operations.   The first hopsital ($i=1$) is the Bristol Royal Infirmary.  \shortciteN{marshall+s07} consider a random effects model of the form
$y_i\sim \text{Binomial}(n_i,p_i)$ where $\log (p_i/(1-p_i))=\beta+u_i$ and $u_i\sim N(0,D)$ so that $u_i$ are hospital specific random effects, and 
they consider formal measures of conflict involving the prior for $u_i$ given $D$.   Particular interest is in whether there is a prior data conflict for $i=1$ (Bristol)
which would indicate that this hospital is unusual compared to the others.  In our analysis here we consider priors on $\beta$ and $D$ where $\beta\sim N(0,1000)$ and
$\log D\sim N(-3.5,1)$ which were chosen to be roughly similar to priors chosen in \shortciteN{tan+n14} for this example.  
So we have a hierarchical prior, $g(\theta)=g(u,\beta,D)=g(u|D)g(\beta,D)$ and we can use our methods for checking hierarchical priors to check for conflict involving
each of the $u_i$.  

We will use a multivariate normal variational approximation to $g(\theta|y)$ (but with $D$ transformed by taking logs) and computed
using the method described in \shortciteN{Kucukelbir2016}.  The conditional prior $p(u|D)$ is normal, and in the variational
posterior the conditional for $u$ given $\beta,D$ is also normal, so that conditional prior to (variational) posterior divergences can be computed in closed form.
For checking for conflict for the $u_i$s we will use the statisic $\text{KL}_1(y)=\lim_{\alpha\rightarrow 1}R_{\alpha 1}(y)$, except that we 
replace the conditional posterior and prior for $u$ given $\beta,D$ in the definition (\ref{ralphay2}) with that of $u_i$ given $\beta, D$ when checking $u_i$. 
This is because we are intersted in checking for conflicts for individual hospital specific effects.  We will approximate $\text{KL}_1(y)$ by 
$\text{KL}_1^*(y)$ obtained by replacing all computations involving the true posterior with the equivalent calculations for the variational
Gaussian posterior.  

Figure \ref{posteriors} shows for the observed data the variational posterior distribution, together with the true posterior approximated
by MCMC.  Table~\ref{Table:Bristol} also shows our conflict $p$-values for the different hospitals.  Also listed are cross-validated mixed predictive $p$-values obtained
by the method of \shortciteN{marshall+s07} by MCMC and given in 
\shortciteN[Table 1]{tan+n14}, as well as a cross-validated version of our divergence based $p$-values.  
The cross-validated divergence based $p$-values use the posterior distribution for $(\beta,D)$ obtained when leaving out the $i$th observation,
$g(\theta_2|y_{\text{obs,-i}})$, instead of $g(\theta_2|y_{\text{obs}})$ in the definition of the reference distribution (\ref{klcheck1})
and in taking the expectation in (\ref{Ralpha1}).
We can see that the $p$-values are similar although the priors on the parameters $(\beta,D)$ were not exactly
the same in Tan and Nott's analysis.  For comparison with previous analyses of the data, we have computed a one-sided version of our conflict $p$-value here, 
which makes sense because excess mortality is of interest.  We have modified our $p$-value measuring surprise to
$p_{\text{KL}1}=P\Bigl(\text{KL}_1(Y)\geq \text{KL}_1(y_{\text{obs}})\mbox{ and } E_q(u_i|Y)>0\Bigr)$ for clusters $i$ with $E_q(u_i|y_{\text{obs}})>0$, and to
$p_{\text{KL}1}=P\Bigl(\text{KL}_1(Y)\leq \text{KL}_1(y_{\text{obs}})\Bigr)+P\Bigl(\text{KL}_1(Y)\geq KL_1(y_{\text{obs}})\mbox{ and }E_q(u_i|Y)>0\Bigr)$ for clusters $i$ with $E(u_i|y_{\text{obs}})<0$,
where in these expressions $E_q(\cdot)$ denotes expectation with respect to the appropriate variational posterior distribution.   Although it is not expected that these conflict
$p$-values should be exactly the same, it is seen that they give a similar picture about the degree of consistency of the data for each hospital with the hierarchical prior.  
\begin{figure}[htp]
\begin{center}
\begin{tabular}{cc}
\multicolumn{2}{c}{\includegraphics[height=130mm,width=125mm]{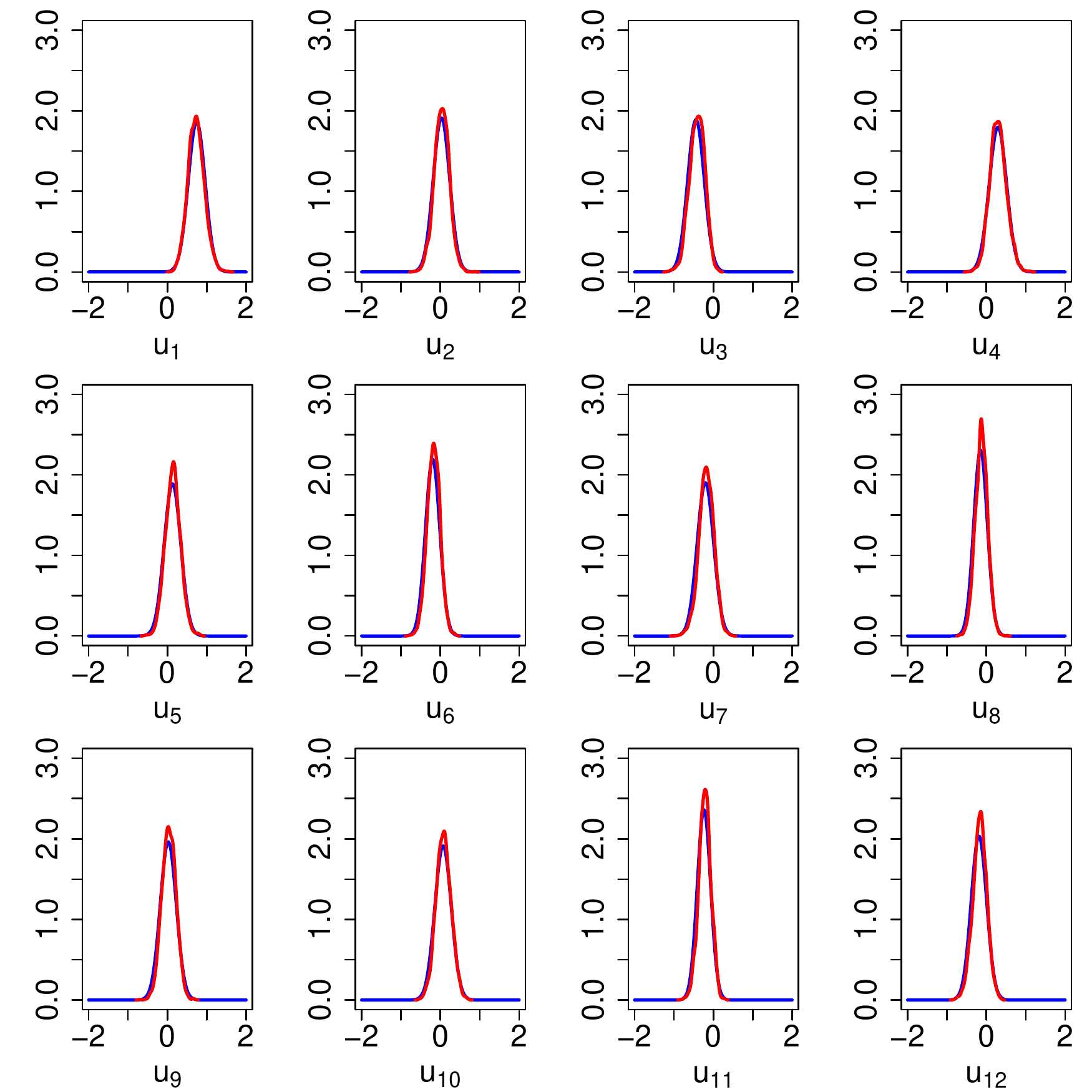}}  \\
\includegraphics[width=60mm]{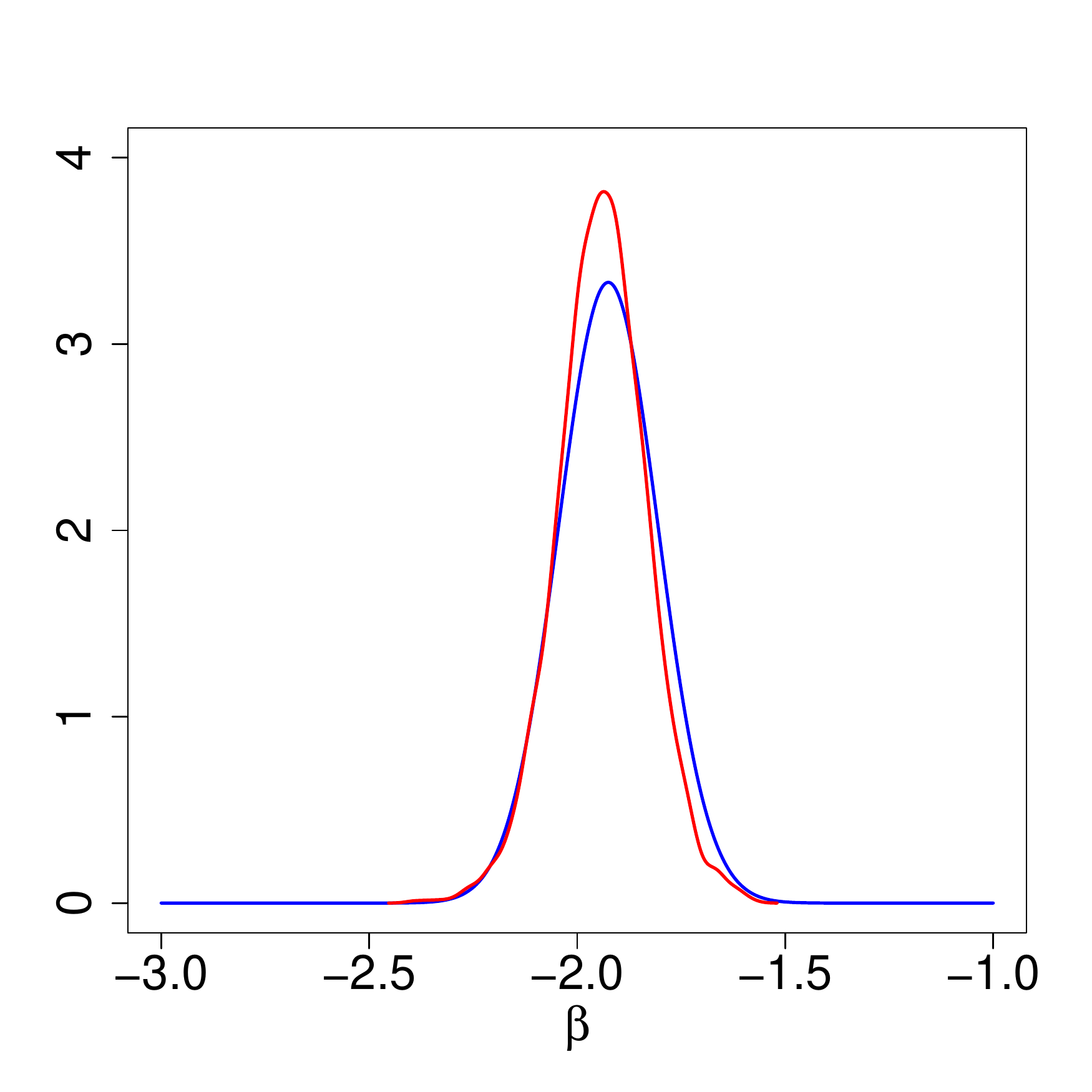} & \includegraphics[width=60mm]{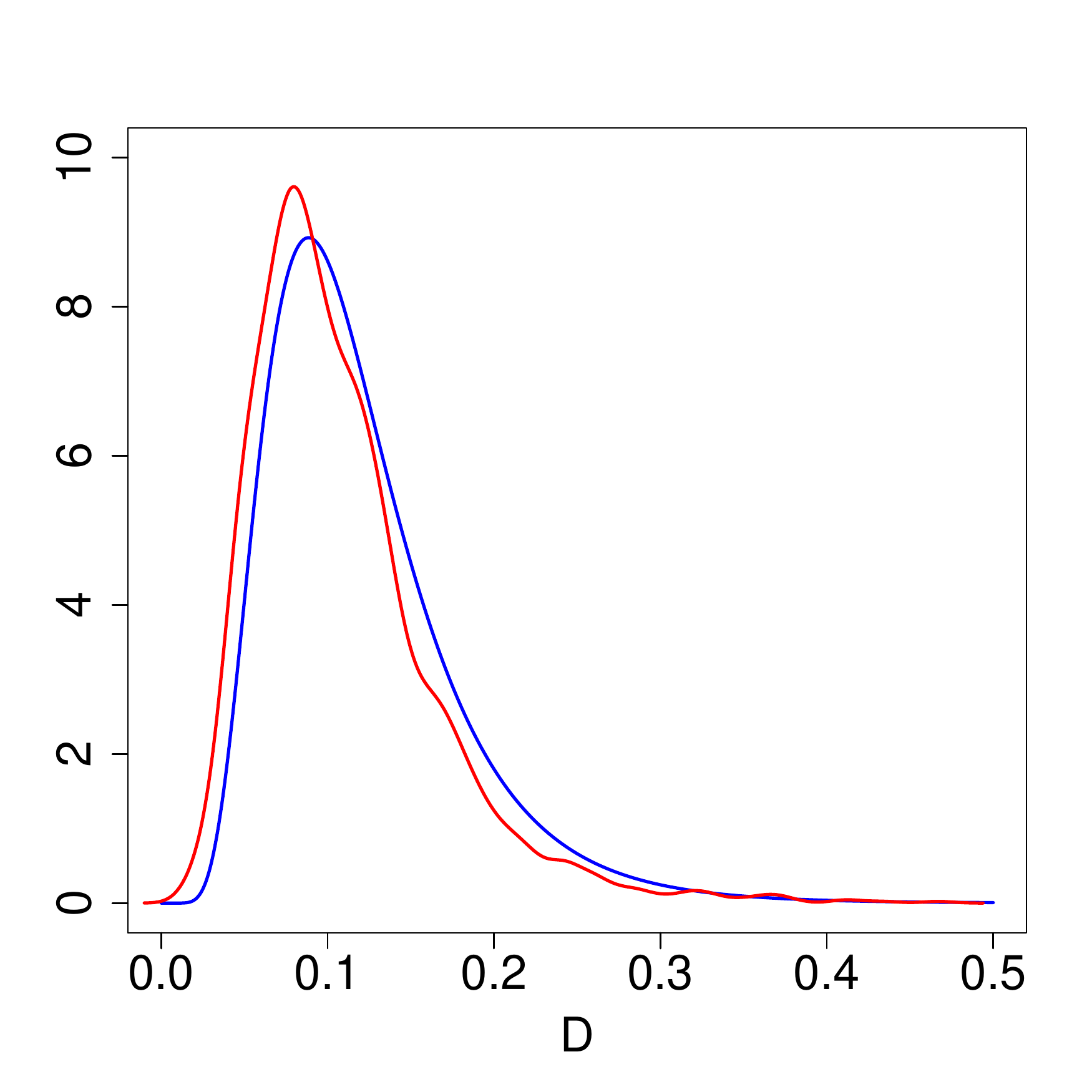} 
\end{tabular}
\end{center}
\caption{\label{posteriors} 
Marginal posterior distributions computed by MCMC (red) and Gaussian variational posteriors (blue) for $u$ (top) and $(\beta,D)$ (bottom).
}
\end{figure}

\begin{table}
\centering
\begin{minipage}[c]{0.6\textwidth}
\caption{\label{Table:Bristol} Cross-validatory conflict $p$-values using the method of Marshall and Spiegelhalter ($p_{\text{MS,CV}}$), KL divergence conflict $p$-values  ($p_{\text{KL}}$), and cross-validated KL divergence $p$-values ($p_{\text{KL,CV}}$) for hospital specific random effects}
\end{minipage}
\begin{center}
\begin{tabular}{cccc}
Hospital & $p_{\text{MS,CV}}$  & $p_{\text{KL}}$ & $p_{\text{KL,CV}}$ \\
\hline 
Bristol & 0.001 & 0.010 & 0.002 \\
Leicester & 0.436 & 0.527 & 0.516 \\
Leeds & 0.935 & 0.912 & 0.947 \\
Oxford & 0.125 & 0.173 & 0.123 \\
Guys & 0.298 & 0.398 & 0.383 \\
Liverpool & 0.720 & 0.690 & 0.745\\
Southampton & 0.737 & 0.680 & 0.715\\
Great Ormond St & 0.661 & 0.595 & 0.628\\
Newcastle & 0.440 & 0.455 & 0.430 \\
Harefield & 0.380 & 0.474 & 0.452 \\
Birmingham & 0.763 & 0.761 & 0.787\\
Brompton & 0.721 & 0.591 & 0.631 \\
\hline
\end{tabular}
\end{center}
\end{table}
\end{exmp}

\section{Discussion}

We have proposed a new approach for prior-data conflict assessment based on comparing the prior to posterior R\'{e}nyi divergence to its distribution
under the prior predictive for the data.  The method can be extended to hierarchical settings where it is desired to check different components of a prior
distribution, and has some interesting connections with the methodology of \shortciteN{evans+m06} and with Jeffreys' and reference prior distributions.  
It works well in the examples we have examined, and we have suggested the use of variational approximations for making the methodology
implementable in complex settings.

There are a number of ways that this work could be further developed.    One line of future development concerns the computational approximations developed
in Section 5, which can no doubt be improved.  On the more statistical side, \shortciteN{evans+j11} define a notion of weak informativity of 
a prior with respect to a given base prior, inspired by ideas of \shortciteN{gelman06}, and their particular formulation of this concept makes use of the notion of
prior-data conflict checks.  It will be interesting to examine how the prior-data conflict checks we have developed here perform in relation to this application.

\subsection*{Acknowledgements}

David Nott was supported by a Singapore Ministry of Education Academic Research
Fund Tier 2 grant (R-155-000-143-112).  Bergthold-Georg Englert's work is funded by the Singapore Ministry of Education (partly 
through the Academic Research Fund Tier 3 MOE2012-T3-1-009) and the National Research Foundation of Singapore.
Michael Evans's work was supported by a Natural Sciences and Engineering Research Council of Canada Grant Number 10671.

\bibliographystyle{chicago}
\bibliography{conflict-KL}

\begin{thebibliography}{}

\bibitem[\protect\citeauthoryear{{Al-Labadi} and {Evans}}{{Al-Labadi} and
  {Evans}}{2015}]{allabadi+e15}
{Al-Labadi}, L. and M.~{Evans} (2015).
\newblock {Optimal robustness results for some Bayesian procedures and the
  relationship to prior-data conflict. arXiv1504.06898}.

\bibitem[\protect\citeauthoryear{Albert}{Albert}{2009}]{albert2009bayesian}
Albert, J. (2009).
\newblock {\em Bayesian computation with R}.
\newblock Springer Science \& Business Media.

\bibitem[\protect\citeauthoryear{Andrade and O'Hagan}{Andrade and
  O'Hagan}{2006}]{andrade+o06}
Andrade, J. A.~A. and A.~O'Hagan (2006).
\newblock Bayesian robustness modeling using regularly varying distributions.
\newblock {\em Bayesian Analysis\/}~{\em 1}, 169--188.

\bibitem[\protect\citeauthoryear{Attias}{Attias}{1999}]{attias99}
Attias, H. (1999).
\newblock Inferring parameters and structure of latent variable models by
  variational {B}ayes.
\newblock In K.~Laskey and H.~Prade (Eds.), {\em Proceedings of the 15th
  Conference on Uncertainty in Artificial Intelligence}, San Francisco, CA,
  pp.\  21--30. Morgan Kaufmann.

\bibitem[\protect\citeauthoryear{Baskurt and Evans}{Baskurt and
  Evans}{2013}]{baskurt+e13}
Baskurt, Z. and M.~Evans (2013).
\newblock Hypothesis assessment and inequalities for {B}ayes factors and
  relative belief ratios.
\newblock {\em Bayesian Analysis\/}~{\em 8\/}(3), 569--590.

\bibitem[\protect\citeauthoryear{Bayarri and Berger}{Bayarri and
  Berger}{2000}]{bayarri+b00}
Bayarri, M.~J. and J.~O. Berger (2000).
\newblock P values for composite null models (with discussion).
\newblock {\em Journal of the American Statistical Association\/}~{\em 95}, pp.
  1127--1142.

\bibitem[\protect\citeauthoryear{Bayarri and Castellanos}{Bayarri and
  Castellanos}{2007}]{bayarri+c07}
Bayarri, M.~J. and M.~E. Castellanos (2007).
\newblock Bayesian checking of the second levels of hierarchical models.
\newblock {\em Statistical Science\/}~{\em 22}, 322--343.

\bibitem[\protect\citeauthoryear{Berger, Bernardo, and Sun}{Berger
  et~al.}{2009}]{berger+bs09}
Berger, J.~O., J.~M. Bernardo, and D.~Sun (2009).
\newblock The formal definition of reference priors.
\newblock {\em The Annals of Statistics\/}~{\em 37\/}(2), 905--938.

\bibitem[\protect\citeauthoryear{Bousquet}{Bousquet}{2008}]{bousquet08}
Bousquet, N. (2008).
\newblock Diagnostics of prior-data agreement in applied {B}ayesian analysis.
\newblock {\em Journal of Applied Statisics\/}~{\em 35}, 1011--1029.

\bibitem[\protect\citeauthoryear{Box}{Box}{1980}]{box80}
Box, G. E.~P. (1980).
\newblock {Sampling and Bayes' inference in scientific modelling and robustness
  (with discussion)}.
\newblock {\em Journal of the Royal Statistical Society, Series A\/}~{\em 143},
  383--430.

\bibitem[\protect\citeauthoryear{Carota, Parmigiani, and Polson}{Carota
  et~al.}{1996}]{carota+pp96}
Carota, C., G.~Parmigiani, and N.~G. Polson (1996).
\newblock Diagnostic measures for model criticism.
\newblock {\em Journal of the American Statistical Association\/}~{\em
  91\/}(434), 753--762.

\bibitem[\protect\citeauthoryear{Dahl, G\r{a}semyr, and Natvig}{Dahl
  et~al.}{2007}]{dahl+gn07}
Dahl, F.~A., J.~G\r{a}semyr, and B.~Natvig (2007).
\newblock A robust conflict measure of inconsistencies in {B}ayesian
  hierarchical models.
\newblock {\em Scandinavian Journal of Statistics\/}~{\em 34}, 816--828.

\bibitem[\protect\citeauthoryear{Dey, Gelfand, Swartz, and Vlachos}{Dey
  et~al.}{1998}]{dey+gsv98}
Dey, D.~K., A.~E. Gelfand, T.~B. Swartz, and P.~K. Vlachos (1998).
\newblock A simulation-intensive approach for checking hierarchical models.
\newblock {\em Test\/}~{\em 7}, 325--346.

\bibitem[\protect\citeauthoryear{Evans}{Evans}{2015}]{evans15}
Evans, M. (2015).
\newblock {\em Measuring Statistical Evidence Using Relative Belief}.
\newblock Taylor \& Francis.

\bibitem[\protect\citeauthoryear{Evans and Jang}{Evans and
  Jang}{2010}]{evans+j10}
Evans, M. and G.~H. Jang (2010).
\newblock Invariant p-values for model checking.
\newblock {\em The Annals of Statistics\/}~{\em 38}, 512--525.

\bibitem[\protect\citeauthoryear{Evans and Jang}{Evans and
  Jang}{2011a}]{evans+j11b}
Evans, M. and G.~H. Jang (2011a).
\newblock A limit result for the prior predictive applied to checking for
  prior-data conflict.
\newblock {\em Statistics and Probability Letters\/}~{\em 81\/}(8), 1034 --
  1038.

\bibitem[\protect\citeauthoryear{Evans and Jang}{Evans and
  Jang}{2011b}]{evans+j11}
Evans, M. and G.~H. Jang (2011b).
\newblock Weak informativity and the information in one prior relative to
  another.
\newblock {\em Statistical Science\/}~{\em 26}, 423--439.

\bibitem[\protect\citeauthoryear{Evans and Moshonov}{Evans and
  Moshonov}{2006}]{evans+m06}
Evans, M. and H.~Moshonov (2006).
\newblock Checking for prior-data conflict.
\newblock {\em Bayesian Analysis\/}~{\em 1}, 893--914.

\bibitem[\protect\citeauthoryear{Gelman}{Gelman}{2006}]{gelman06}
Gelman, A. (2006).
\newblock Prior distributions for variance parameters in hierarchical models.
\newblock {\em Bayesian Analysis\/}~{\em 1}, 1--19.

\bibitem[\protect\citeauthoryear{Gelman, Meng, and Stern}{Gelman
  et~al.}{1996}]{gelman+ms96}
Gelman, A., X.-L. Meng, and H.~Stern (1996).
\newblock Posterior predictive assessment of model fitness via realized
  discrepancies.
\newblock {\em Statistica Sinica\/}~{\em 6}, 733--807.

\bibitem[\protect\citeauthoryear{Gershman, Hoffman, and Blei}{Gershman
  et~al.}{2012}]{gershman+hb12}
Gershman, S., M.~D. Hoffman, and D.~M. Blei (2012).
\newblock Nonparametric variational inference.
\newblock In {\em Proceedings of the 29th International Conference on Machine
  Learning, {ICML} 2012}.

\bibitem[\protect\citeauthoryear{Ghosh, Delampady, and Samanta}{Ghosh
  et~al.}{2006}]{ghosh+ds06}
Ghosh, J., M.~Delampady, and T.~Samanta (2006).
\newblock {\em An Introduction to Bayesian Analysis: Theory and Methods}.
\newblock Springer Texts in Statistics. Springer New York.

\bibitem[\protect\citeauthoryear{Ghosh}{Ghosh}{2011}]{ghosh11}
Ghosh, M. (2011).
\newblock Objective priors: An introduction for frequentists.
\newblock {\em Statistical Science\/}~{\em 26}, 187--202.

\bibitem[\protect\citeauthoryear{Gil, Alajaji, and Linder}{Gil
  et~al.}{2013}]{gil+al13}
Gil, M., F.~Alajaji, and T.~Linder (2013).
\newblock R{\'{e}}nyi divergence measures for commonly used univariate
  continuous distributions.
\newblock {\em Information Sciences\/}~{\em 249}, 124--131.

\bibitem[\protect\citeauthoryear{G\r{a}semyr and Natvig}{G\r{a}semyr and
  Natvig}{2009}]{gasemyr+n09}
G\r{a}semyr, J. and B.~Natvig (2009).
\newblock Extensions of a conflict measure of inconsistencies in {B}ayesian
  hierarchical models.
\newblock {\em Scandinavian Journal of Statistics\/}~{\em 36}, 822--838.

\bibitem[\protect\citeauthoryear{Guttman}{Guttman}{1967}]{guttman67}
Guttman, I. (1967).
\newblock The use of the concept of a future observation in goodness-of-fit
  problems.
\newblock {\em Journal of the Royal Statistical Society, Series B\/}~{\em 29},
  83--100.

\bibitem[\protect\citeauthoryear{Hershey and Olsen}{Hershey and
  Olsen}{2007}]{hershey+o07}
Hershey, J.~R. and P.~A. Olsen (2007).
\newblock Approximating the {K}ullback {L}eibler divergence between {G}aussian
  mixture models.
\newblock In {\em 2007 IEEE International Conference on Acoustics, Speech and
  Signal Processing - ICASSP '07}, Volume~4, pp.\  IV--317--IV--320.

\bibitem[\protect\citeauthoryear{Hjort, Dahl, and Steinbakk}{Hjort
  et~al.}{2006}]{hjort+ds06}
Hjort, N.~L., F.~A. Dahl, and G.~H. Steinbakk (2006).
\newblock Post-processing posterior predictive p-values.
\newblock {\em Journal of the American Statistical Association\/}~{\em 101},
  1157--1174.

\bibitem[\protect\citeauthoryear{Jaynes}{Jaynes}{1976}]{jaynes76}
Jaynes, E. (1976).
\newblock {\em Confidence Intervals vs Bayesian Intervals (1976)}, pp.\
  175--267.
\newblock Dordrecht: Reidel Publishing Company.

\bibitem[\protect\citeauthoryear{Johnson}{Johnson}{1970}]{johnson70}
Johnson, R.~A. (1970).
\newblock Asymptotic expansions associated with posterior distributions.
\newblock {\em The Annals of Mathematical Statistics\/}~{\em 41\/}(3),
  851--864.

\bibitem[\protect\citeauthoryear{Jordan, Ghahramani, Jaakkola, and Saul}{Jordan
  et~al.}{1999}]{jordan+gjs99}
Jordan, M.~I., Z.~Ghahramani, T.~S. Jaakkola, and L.~K. Saul (1999).
\newblock An introduction to variational methods for graphical models.
\newblock {\em Machine Learning\/}~{\em 37}, 183--233.

\bibitem[\protect\citeauthoryear{Kucukelbir, Tran, Ranganath, Gelman, and
  Blei}{Kucukelbir et~al.}{2016}]{Kucukelbir2016}
Kucukelbir, A., D.~Tran, R.~Ranganath, A.~Gelman, and D.~M. Blei (2016).
\newblock Automatic differentiation variational inference.
\newblock arXiv: 1603.00788.

\bibitem[\protect\citeauthoryear{Li, Shang, Ng, and Englert}{Li
  et~al.}{2016}]{li+sne16}
Li, X., J.~Shang, H.~K. Ng, and B.-G. Englert (2016).
\newblock {Optimal error intervals for properties of the quantum state.
  arXiv1602.05780}.

\bibitem[\protect\citeauthoryear{Liese and Vajda}{Liese and
  Vajda}{1987}]{liese+v87}
Liese, F. and I.~Vajda (1987).
\newblock {\em Convex Statistical Distances}.
\newblock Teubner-Texte zur Mathematik. Teubner.

\bibitem[\protect\citeauthoryear{Marshall and Spiegelhalter}{Marshall and
  Spiegelhalter}{2007}]{marshall+s07}
Marshall, E.~C. and D.~J. Spiegelhalter (2007).
\newblock Identifying outliers in {B}ayesian hierarchical models: a
  simulation-based approach.
\newblock {\em Bayesian Analysis\/}~{\em 2}, 409--444.

\bibitem[\protect\citeauthoryear{O'Hagan}{O'Hagan}{2003}]{ohagan03}
O'Hagan, A. (2003).
\newblock {HSS} model criticism (with discussion).
\newblock In P.~J. Green, N.~L. Hjort, and S.~T. Richardson (Eds.), {\em Highly
  Structured Stochastic Systems}, pp.\  423--453. Oxford University Press.

\bibitem[\protect\citeauthoryear{Presanis, Ohlssen, Spiegelhalter, and
  Angelis}{Presanis et~al.}{2013}]{presanis+osd13}
Presanis, A.~M., D.~Ohlssen, D.~J. Spiegelhalter, and D.~D. Angelis (2013).
\newblock Conflict diagnostics in directed acyclic graphs, with applications in
  {B}ayesian evidence synthesis.
\newblock {\em Statistical Science\/}~{\em 28}, 376--397.

\bibitem[\protect\citeauthoryear{{Reimherr}, {Meng}, and {Nicolae}}{{Reimherr}
  et~al.}{2014}]{reimherr+mn14}
{Reimherr}, M., X.-L. {Meng}, and D.~L. {Nicolae} (2014).
\newblock {Being an informed Bayesian: Assessing prior informativeness and
  prior likelihood conflict. arXiv1406.5958}.

\bibitem[\protect\citeauthoryear{R\'{e}nyi}{R\'{e}nyi}{1961}]{renyi61}
R\'{e}nyi, A. (1961).
\newblock On measures of entropy and information.
\newblock In {\em Proceedings of the Fourth Berkeley Symposium on Mathematical
  Statistics and Probability, Volume 1: Contributions to the Theory of
  Statistics}, Berkeley, Calif., pp.\  547--561. University of California
  Press.

\bibitem[\protect\citeauthoryear{Robins, van~der Vaart, and Ventura}{Robins
  et~al.}{2000}]{robins+vv00}
Robins, J.~M., A.~van~der Vaart, and V.~Ventura (2000).
\newblock Asymptotic distribution of $p$-values in composite null models.
\newblock {\em Journal of the American Statistical Association\/}~{\em 95},
  1143--1156.

\bibitem[\protect\citeauthoryear{Rohde and Wand}{Rohde and
  Wand}{2015}]{rohde+w15}
Rohde, D. and M.~P. Wand (2015).
\newblock Mean field variational {B}ayes: general principles and numerical
  issues.
\newblock {https://works.bepress.com/matt_wand/15/}.

\bibitem[\protect\citeauthoryear{Rubin}{Rubin}{1984}]{rubin84}
Rubin, D.~B. (1984).
\newblock Bayesianly justifiable and relevant frequency calculations for the
  applied statistician.
\newblock {\em Annals of Statistics\/}~{\em 12}, 1151--1172.

\bibitem[\protect\citeauthoryear{Salimans and Knowles}{Salimans and
  Knowles}{2013}]{salimans+k13}
Salimans, T. and D.~A. Knowles (2013).
\newblock Fixed-form variational posterior approximation through stochastic
  linear regression.
\newblock {\em Bayesian Analysis\/}~{\em 8}, 837--882.

\bibitem[\protect\citeauthoryear{Scheel, Green, and Rougier}{Scheel
  et~al.}{2011}]{scheel+gr11}
Scheel, I., P.~J. Green, and J.~C. Rougier (2011).
\newblock A graphical diagnostic for identifying influential model choices in
  {B}ayesian hierarchical models.
\newblock {\em Scandinavian Journal of Statistics\/}~{\em 38\/}(3), 529--550.

\bibitem[\protect\citeauthoryear{Spiegelhalter, Aylin, Best, Evans, and
  Murray}{Spiegelhalter et~al.}{2002}]{spiegelhalter+abem02}
Spiegelhalter, D., P.~Aylin, N.~Best, S.~Evans, and G.~Murray (2002).
\newblock Commissioned analysis of surgical performance using routine data:
  lessons from the bristol inquiry.
\newblock {\em Journal of the Royal Statistical Society, Series A\/}~{\em 165},
  191--221.

\bibitem[\protect\citeauthoryear{Tan and Nott}{Tan and Nott}{2014}]{tan+n14}
Tan, L. S.~L. and D.~J. Nott (2014).
\newblock A stochastic variational framework for fitting and diagnosing
  generalized linear mixed models.
\newblock {\em {B}ayesian Analysis\/}~{\em 9}, 963--1004.

\bibitem[\protect\citeauthoryear{Winn and Bishop}{Winn and
  Bishop}{2005}]{winn+b05}
Winn, J. and C.~M. Bishop (2005).
\newblock Variational message passing.
\newblock {\em Journal of Machine Learning Research\/}~{\em 6}, 661--694.

\end{thebibliography}

\end{document}